\documentstyle[aaspp4]{article}
\newcommand{\bzeta}{\mbox{\boldmath $\zeta$}}
\newcommand{\ratmas}{M_{{\mathit NS}}/M_{{\mathit BH}}}
\newcommand{\ns}{{\mathit NS}}
\newcommand{\bh}{{\mathit BH}}
\newcommand{\inin}{{\mathrm in}}
\newcommand{\ouou}{{\mathrm out}}
\newcommand{\Jseq}{\bar{J}(\bar{d}_G)}
\newcommand{\Eseq}{\bar{E}(\bar{d}_G)}
\newcommand{\Oseq}{\bar{\Omega}(\bar{J})}
\newcommand{\VC}{{\mathit VC}}

\lefthead{K\=oji Ury\=u and Yoshiharu Eriguchi}
\righthead{Stationary structures of irrotational binary systems}

\begin{document}

\title
{Stationary structures of irrotational binary systems --
models for close binary systems of compact stars}
\author
{K\=oji Ury\=u\altaffilmark{1,2,3} and Yoshiharu Eriguchi
\altaffilmark{3}}
\affil
{$^1$International Center for Theoretical Physics, 
Strada Costiera 11, Trieste 34100, Italy}
\affil
{$^2$SISSA, Via Beirut 2-4, Trieste 34013, Italy}
\affil
{$^3$Department of Earth Science and Astronomy,
Graduate School of Arts and Sciences,
University of Tokyo, Komaba, Meguro, Tokyo 153, Japan}
\authoremail{uryu@ictp.trieste.it}
\authoremail{eriguchi@valis.c.u-tokyo.ac.jp}
\begin{abstract}
We propose a new numerical method to calculate irrotational binary systems
composed of compressible gaseous stars in Newtonian gravity.  Assuming 
irrotationality, i.e. vanishing of the vorticity vector everywhere in the 
star in the inertial frame, we can introduce the velocity potential for the 
flow field.  Using this velocity potential we can derive a set of 
basic equations for stationary states which consist of (i) the generalized 
Bernoulli equation, (ii) the Poisson equation for the Newtonian 
gravitational potential and (iii) the equation for the velocity potential 
with the Neumann type boundary condition.  We succeeded in developing a
new code to compute numerically exact solutions to these equations for the 
first time.  Such irrotational configurations of binary systems are  
appropriate models for realistic neutron star binaries composed of 
inviscid gases, just prior to coalescence of two stars caused by emission 
of gravitational waves.  Accuracies of our numerical solutions are so high
that we can compute reliable models for fully deformed final stationary 
configurations and hence determine the inner most stable circular orbit of 
binary neutron star systems under the approximations of weak gravity and 
inviscid limit.  
\end{abstract}

\section{Introduction}

Historically, many authors have tried to obtain equilibrium 
configurations of self-gravitating close binary systems.  The classical 
attempts to solve this problem had been made by constructing approximate 
solutions for binary systems.  As is well known, the first quantitative 
approach was made by Roche who treated a synchronously rotating 
incompressible fluid around a rigid sphere.  By including appropriate terms 
of the tidal potential from the sphere, the stationary state of the fluid 
can be approximated by an ellipsoidal configuration. In order to treat more 
realistic situations, synchronously rotating incompressible fluid-fluid 
binary systems were investigated by Darwin who also used the ellipsoidal 
approximation (see e.g. \cite{ch69}).  For non-synchronously rotating 
binary systems, Aizenman~(1968) studied incompressible fluid binary systems 
with internal motion also by using the ellipsoidal approximation.  
The situation has been changing for these 20 years.  Not only 
incompressible binary systems but also compressible binary configurations 
have been able to be constructed without using the ellipsoidal approximation, 
because several powerful numerical schemes to obtain deformed self-gravitating 
stars have been developed (see e.g. \cite{ehs82} ; \cite{eh83} ; 
\cite{he84a} ; \cite{he84b} ; \cite{ha86}).  

Apart from these developments for solving binary star systems, the problem 
of obtaining stationary states of close binary star 
systems as exactly as possible has, recently, become one of the important 
issues in relativistic astrophysics because they will provide models for 
binary neutron star systems just prior to coalescence due to emission of 
gravitational waves (hereafter GW).  Outcomes of such merging processes of 
binary neutron star systems may be possible sources of astrophysically 
important phenomena such as $\gamma$-ray bursts (see e.g. \cite{pa86}).  
Furthermore, coalescing binary systems are the promising sources of GW 
which may be detected by the ground based interferometric detectors of 
GW in the early stage of the next century 
(LIGO/VIRGO/TAMA/GEO, see e.g. \cite{ab92} ; \cite{th94}).  

The evolution of such close binary neutron star systems due to GW emission 
can be approximated well by quasi-stationary states until just before the 
coalescence stage, since the time scale in which the binary separation 
decreases is longer than the orbital period (see e.g. \cite{st83}).  
The final separation of two component stars where dynamical merging process 
starts is called an inner most stable circular orbit (ISCO) of binary neutron 
stars (see \cite{lrs93a} and references therein).  The ISCO indicates 
implicitly the upper limit of the radius of neutron stars and hence gives 
some information about the equation of state of the neutron matter.  
Consequently, in these several years, relativistic astrophysicists have 
tried to solve equilibrium configurations of highly deformed close binary 
neutron star systems by devising numerical schemes for the binaries.  
Theoretically obtained results for the ISCO is to be compared with the 
observational data of GW detectors in the future, which will give important 
information about the internal structure of neutron stars (\cite{sh97} ; 
\cite{ba97a}, 1997b, 1997c ; see also \cite{bgm97} for a review). 

The scenario of the final evolution of close binary neutron star systems
was considered by Kochanek~(1992) and Bildsten \& Cutler~(1992).  They 
pointed out that flow fields inside the component stars of a binary system 
just prior to coalescence will be {\it irrotational}, i.e. the vorticity 
seen from the inertial frame vanishes just prior to coalescence.  According 
to their results, irrotationality is realized because the viscosity of the
neutron matter is too weak to synchronize the spin and the orbital angular 
velocity during the evolution just before merging.  It is precisely this 
stage of the binary evolution on which we will be focussing in the present 
paper and, for this, the approximation of irrotationality seems likely to 
be very good.  We will explain the irrotational state in the next section.  

In general, to construct stationary configurations of irrotational binary 
systems, it is necessary to treat internal flows in the rotating frame of 
reference in which the stellar figure is seen to be fixed, although there 
is fluid motion within the fixed figure.  However, in most cases, it is 
very difficult to handle generic internal flows or spins of stationary 
compressible stars not only for binary systems but also even for single 
stars.  Consequently, for compressible binary stars, numerical computations 
of fully deformed configurations have been carried out only under the 
assumption that binary stars are rotating rigidly and synchronously or 
a semi-analytic method by using the ellipsoidal approximation has been 
employed.  In particular, Lai, Rasio \& Shapiro~(1993b) developed the 
variational method in which the ellipsoidal approximation is used (LRS1 
hereafter, see also \cite{lrs97} for the post-Newtonian case).  The exact 
treatment of generic flows in such non-axisymmetric 
configurations has only been discussed by the present authors for extended 
models of Dedekind-like configurations for compressible single 
stars~(\cite{ue96} and the references therein).  

However, for irrotational binary configurations, we recently succeeded 
in formulating a new scheme to solve several types of new stationary 
states of irrotational binary systems in Newtonian gravity (\cite{ue98a}, 
hereafter Paper I).  We could develop that formulation by introducing 
the velocity potential for the flow field.  The new scheme is not only 
the first computation of binary configurations which are rotating 
non-synchronously but also an important development to construct models 
for realistic neutron stars in the inviscid limit.  In this paper we will 
explain the computational scheme and discuss the results in detail.  

This paper is organized as follows.  In section 2, we discuss the 
assumption of irrotationality, derive the basic equations, and present 
boundary conditions.  In section 3, we describe specific techniques used 
in our actual numerical computations: choice of coordinate systems, 
introduction of the surface fitted coordinate, non-dimensional variables 
and choice of physical parameters.  The most delicate point in solving 
the irrotational configurations is related to the way how to treat the 
elliptic partial differential equation with the Neumann type boundary 
condition for the velocity potential.  We introduce the solving method of 
the velocity potential in detail in this section.  We also explain 
the iteration procedure.  

In section 4, computational results of our new scheme will 
be presented. In particular, we tabulate the results for the irrotational 
gaseous star--gaseous star binary systems with equal mass.  These 
solutions correspond to so-called Irrotational Darwin-Riemann 
(IDR hereafter) configurations for compressible gases.  We also show 
the results for point source--irrotational gaseous star binary systems 
with various mass ratios.  These are so-called Irrotational Roche-Riemann 
(IRR hereafter) configurations for compressible stars.  The latter system 
can be considered to mimic configurations of black hole--neutron star 
systems.  We compare our results with those obtained under the ellipsoidal 
approximation by Lai, Rasio \& Shapiro (hereafter LRS) in a series of 
papers (\cite{lrs93b} (LRS1), 1994a hereafter LRS2, 1994b) and also with 
our recent results for the irrotational solutions of {\it incompressible} 
binary systems which have been computed by a totally different 
method from the present one (\cite{ue98b}, hereafter Paper II).  Since 
the new scheme turns out to be sufficiently accurate from this comparison, 
our solutions can be treated as reliable ones even for highly deformed 
equilibrium configurations of compressible irrotational binary systems.  
By using sequences of stationary solutions, the dynamical stability of binary 
systems can be discussed and the final fate of binary neutron star systems 
can be clarified.  This is related to the determination of the ISCO for 
the binary systems as mentioned before.  In section 5, we summarize our 
computational results and discuss the remaining problems about the 
realistic close binary neutron star systems.  

Our present results are not only extending the results of the classical 
problem of equilibrium configurations of self-gravitating gases but also 
providing new realistic models of binary neutron star systems such as
a neutron star - neutron star system, or a black hole - neutron star system.  
In such applications, we can regard the gaseous stars as models
for neutron stars (NS) and the point sources as models for black holes (BH). 
%
%
%
%
%

\section{Formulation for irrotational configurations}

As discussed in Introduction, we will treat stationary states of {\it inviscid}
and {\it irrotational} gaseous binary star systems in this paper.  Since the 
binary configurations are non-axisymmetric in nature, the system cannot be
stationary in the inertial frame. However, configurations can be in stationary 
states if seen from a certain rotating frame whose angular velocity
is that of the orbital motion of the binary star system.  First of all
we will explain the validity of the irrotational assumption.

\subsection{Validity of irrotationality for binary neutron star systems}

As mentioned in Introduction, Kochanek~(1992) pointed out that the velocity 
field of the component inviscid star of a binary system just prior to 
coalescence becomes {\it irrotational}.  This can be explained as follows.
For inviscid gases,  Ertel's theorem holds: the ratio of the vorticity 
vector in the inertial frame to the density of a fluid element, 
$\bzeta_0/\rho$, is conserved even under the existence of a potential force 
such as the gravitational radiation reaction (\cite{mi74}).  Here $\rho$
is the density and the vorticity vector in the inertial frame, $\bzeta_0$, 
is defined as
\begin{equation}\label{defvi}
\bzeta_0({\bf r}) \equiv \nabla\times{\bf v}({\bf r})\, ,
\end{equation}
where ${\bf r}$ and ${\bf v}({\bf r})$ are the position vector of 
the fluid element and the velocity field seen from the inertial frame, 
respectively.  In the rotating frame in which the binary system can be 
seen in a quasi-stationary state as mentioned above, the value of $\bzeta_0$ 
can be regarded as negligibly small from the following reason.  The vorticity 
vector $\bzeta_0$ can be expressed by using the vorticity vector in the 
rotating frame $\bzeta$ and the orbital angular velocity vector ${\bf \Omega}$ 
as follows:
\begin{equation}\label{vorome}
\bzeta_0({\bf r})\,=\,\bzeta({\bf r})\,+\,2{\bf \Omega}\ , 
\end{equation}
where ${\bf \Omega}$ is defined as 
${\bf \Omega} \, = \, \Omega \, {\bf e}_z$,   
${\bf e}_i$ ($i \, = \, x,y,z$) denotes the unit basis vector 
along the coordinate axis, $z$ is the direction parallel to the 
rotational axis of the orbital motion and $\Omega$ is the constant 
orbital angular velocity.  Here the vorticity vector $\bzeta$ is also 
expressed by using the velocity vector of the fluid in the rotational frame, 
${\bf u}({\bf r})$, by 
\begin{equation}\label{defvr}
\bzeta({\bf r}) \equiv \nabla\times{\bf u}({\bf r}) \ .
\end{equation}
The above relation is derived by taking the curl of the following relation 
between the fluid velocity in the inertial frame and that in the rotational 
frame, 
\begin{equation}\label{relv}
{\bf v}({\bf r})\,=\,{\bf u}({\bf r})\,+{\bf \Omega}\times{\bf r} \ . 
\end{equation}

Since the value of $\Omega$ at a close binary state is sufficiently larger 
than that of $\Omega$ and that of $\bzeta$ at a detached phase, and since 
$\bzeta_0$ is conserved during evolution of a binary system, the contribution 
of $\bzeta_0$ to the final values of $\Omega$ and $\bzeta$ can be totally 
negligible.  Therefore we can consider that the {\it realistic} close binary 
neutron star system originated from emission of GW is 
composed of {\it irrotational} gases, i.e. 
\begin{equation}\label{vfree}
\bzeta_0({\bf r})\,=\,0\ , 
\end{equation}
everywhere.  

Accordingly, for irrotational gases, we can assume the existence of the 
velocity potential, $\Phi({\bf r})$, which satisfies the following
relation in the inertial frame:
\begin{equation}\label{defvp}
{\bf v}({\bf r}) \equiv  \nabla\Phi({\bf r}) \ .  
\end{equation}

\subsection{Basic equations and boundary conditions}

Since we will treat inviscid stars in {\it Newtonian} gravity, basic equations 
consist of the Euler equation, the equation of continuity and the 
Poisson equation as follows: 
\begin{equation} \label{euler}
{\partial \, {\bf v} \over \partial \, t}\, + \,{\bf v}\cdot\nabla{\bf v}
\,=\, -{1 \over \rho}\nabla p\,-\,\nabla \phi \ , 
\end{equation}
\begin{equation} \label{conti}
{\partial \, \rho \over \partial \, t} \, + \, 
\nabla \cdot(\rho\,{\bf v})\, = \, 0\ , 
\end{equation}
\begin{equation} \label{poiso}
\triangle \phi\, =\,4 \pi G \rho\ , 
\end{equation}
where  $p$, $\phi$ and $G$ are the pressure, the gravitational 
potential and the gravitational constant, respectively.

For irrotational gases, the integrability condition for the equation
(\ref{euler}) requires that the pressure must be a function of the density,
i.e. that the barotropic relation as follows must be satisfied:
\begin{equation} \label{baror}
p\,=\,p(\rho)\, \ .
\end{equation}
For barotropes, the Euler equation (\ref{euler}) can be integrated to the 
generalized Bernoulli equation.  In the rotating frame, the generalized 
Bernoulli equation is written as follows: 
\begin{equation} \label{gberno}
{\partial \, \Phi \over \partial \, t} \,-\,({\bf \Omega} \times {\bf r})
\cdot \nabla \, \Phi \,+\, {1 \over 2} \left|\, \nabla \, \Phi \,\right|^2 
\,+\,\int{dp \over \rho} \,+\, \phi \, = \, f(t)\, \ ,
\end{equation}
where $f(t)$ is an arbitrary function of time.  The origin of the position 
vector ${\bf r}$ for the fluid element in the star is chosen at the point 
where the rotational axis intersects the equatorial plane.  
For simplicity, we choose the following polytropic relation
as the equation of state:
\begin{equation} \label{poly}
p = K \rho^{1+1/N} = K \Theta^{N+1} \ ,
\end{equation}
where $\Theta$, $N$ and $K$ are the Emden function which is proportional to 
the enthalpy, the polytropic index and a certain constant, respectively.  
Note that the Emden function defined here is different in normalization
from that usually employed in spherical polytropic stars.  The equation of 
continuity (\ref{conti}) is also rewritten by using the velocity potential 
$\Phi$ as follows: 
\begin{equation} \label{vconti}
{\partial \, \rho \over \partial \, t} \, + \,
\nabla\cdot(\rho\,\nabla\Phi)\, = \, 0 \ .
\end{equation}

Since we assume stationarity of configurations in the rotating frame,
we can set following conditions:
\begin{equation} \label{stat1}
{\partial \, \rho \over \partial \, t} \, \equiv \, 0 \, \quad
{\partial \, \Phi \over \partial \, t} \, \equiv \, 0 \quad
{\mathrm and }\quad f(t) \, = \, C \, = \, {\mathit constant} \ .
\end{equation}
Therefore, we can rewrite equations (\ref{gberno}) and (\ref{vconti}) 
as follows:
\begin{equation} \label{nberno}
\Theta \, = \, {1 \over K\,(N+1)}
\, \left[ \, ({\bf \Omega} \times {\bf r}) \cdot \nabla \, \Phi \, - \,
{1 \over 2}\left|\, \nabla\,\Phi\,\right|^2\,-\,\phi \,+\, C \, \right] \ ,
\end{equation}
and 
\begin{equation} \label{pconti}
\triangle \Phi \, = \, N ({\bf \Omega} \times {\bf r} \, - \,
\nabla \, \Phi) \cdot {\nabla \, \Theta \over \Theta}\ ,  
\end{equation}
where we use the relation (\ref{poly}).  

As for the Poisson equation (\ref{poiso}), we use the integral form as 
follows: 
\begin{equation} \label{intpot}
  \phi({\bf r}) \, = \, - \, G
  \int_{\cal V} {\rho({\bf r}^{'}) \over 
\left|\, {\bf r}-{\bf r}^{'} \,\right| }  d^3 {\bf r}^{'}\ , 
\end{equation}
where the integration is performed over the whole volume of the stars 
${\cal V}$.  
The boundary condition for $\phi$ is automatically included in this 
expression.  For point source models, the gravitational potential of
the point source (BH) is expressed as 
\begin{equation} \label{soupot}
\phi_{\bh}({\bf r}) \, = \, - \,{G M_{\bh} \over 
\left|\, {\bf r}-{\bf r}_{\bh} \,\right| }\ , 
\end{equation}
where $M_{\bh}$ and ${\bf r}_{\bh}$ are the mass and the position vector 
of the point source, respectively.  If we substitute these expressions 
for the gravitational potentials into the equation (\ref{nberno}), 
we obtain two equations for two field variables, the Emden function 
$\Theta({\bf r})$ and the velocity potential $\Phi({\bf r})$.  

Boundary conditions for these two variables are as follows.  The pressure 
and accordingly the Emden function should vanish on the stellar 
surface because of the definition of the surface itself.  Concerning
the velocity potential, the fluid element on the stellar surface should 
flow along it in the rotational frame because of stationarity of 
the configuration.  Geometrically speaking, it means that the fluid velocity 
on the surface in this frame ${\bf u}_S$ is perpendicular to the normal 
of the stellar surface ${\bf n}_S$, where the subscript $S$ denotes the 
stellar surface.  Thus these boundary conditions are written as follows:
\begin{equation}\label{bconp}
\Theta({\bf r}_S) \, = \, 0\ , 
\end{equation}
\begin{equation}\label{bconv}
{\bf n}_S\cdot{\bf u}_S\, =\,
{\bf n}_S\cdot({\nabla \Phi}\,-\,{\bf \Omega}\times {\bf r}_S) 
\, = \, 0\ ,
\end{equation}
where we use the relation ($\ref{relv}$) and we denote the position vectors 
of points on the stellar surface by ${\bf r}_S$.  The former condition 
determines the shape of stellar surface.  As seen from the form of the 
equation of continuity (\ref{pconti}) and from that of the boundary 
condition (\ref{bconv}), the problem is reduced to an elliptic partial 
differential equation with a Neumann type boundary condition.  From the 
standpoint of numerical computations, careful treatment is required to solve 
this kind of equations.  Thus we will explain the solving scheme of that 
equation to some extent.  

\subsection{Integral form of the Neumann problem}

Here we consider the following Neumann problem: 
\begin{equation} \label{neue}
\triangle \Phi \, = \,S({\bf r})\ , 
\end{equation}
\begin{equation} \label{neub}
\left. {\bf n}_S \cdot \nabla \Phi\right|_S\,=\, 
{\bf n}_S \cdot({\bf \Omega} \times {\bf r}_S)\ ,
\end{equation}
where $S({\bf r})$ is defined as
\begin{equation} \label{source}
S({\bf r})\, = \, N ({\bf \Omega} \times {\bf r}\,-\,
\nabla \, \Phi) \cdot {\nabla \, \Theta \over \Theta}\ . 
\end{equation}

As is well known, a solution of the Neumann problem for the elliptical partial
differential equation (\ref{neue}) can be expressed as follows:
\begin{equation} \label{intform}
\Phi({\bf r}) \, = \,-{1\over 4\pi}\int_V 
{S({\bf r'})\over\left|\, {\bf r}-{\bf r}^{'} \,\right|} 
d^3 {\bf r}^{'}\,+\,\chi({\bf r})  \ ,
\end{equation}
where $\chi({\bf r})$ is a regular homogeneous solution to the 
Laplace equation inside the star, i.e.
\begin{equation} \label{homoge}
\triangle\,\chi({\bf r}) \, = \, 0 \ . 
\end{equation}
Note that the integral is performed on $V$ which is a volume of 
one gaseous component star.  
Substituting the above expression to the l.h.s. of the boundary condition 
$(\ref{neub})$, we have the following equation:
\begin{equation} \label{intbou}
-{1\over 4\pi}\int_V 
{\bf n}_S\cdot\left(\nabla{1 \over\left|\, {\bf r}-{\bf r}^{'} \,\right|} 
\right)_S\,S({\bf r'})d^3 {\bf r}^{'}
\,+\,\left.{\bf n}_S \cdot\nabla \chi({\bf r})\right|_S
\,=\,{\bf n}_S \cdot({\bf \Omega} \times {\bf r}_S) \ .
\end{equation}
It should be noted that this is a formal solution because there appears 
the unknown function $\Phi$ in the source term.  In other words, this is an
integral equation for unknowns.  Thus, once the Emden function is given,
the velocity potential can be obtained by solving equations~(\ref{intform}),
(\ref{homoge}) and ({\ref{intbou}).

\section{Implementation for actual computations}

In the previous section, we have written down the basic equations for the 
irrotational binary configurations.  For our actual computations, we use 
equations (\ref{nberno}), (\ref{intform}) and  (\ref{homoge}) and the 
boundary conditions (\ref{bconp}) and (\ref{intbou}).  We will find 
solutions to these equations by using an iterative scheme explained later.
To accomplish a robust iteration scheme to converged solutions, we need 
to devise several techniques which have been obtained after a number
of attempts to solve equilibrium configurations of rotating stars.  
We will describe some techniques in this section: choice of the coordinates, 
choice of parameters and the iteration scheme.  Since the Neumann problems 
have not been treated in the theory of rotating stars except for 
Eriguchi~(1990), we will explain the solving method for it in detail.  

\subsection{Choice of coordinates}

Since we treat the BH--NS binary systems or the equal mass NS--NS binary 
systems, we only need to compute structures of one gaseous component star.  
The Cartesian coordinates $(x,y,z)$ whose origin is located at the 
intersection of the rotational axis and the equatorial plane are used 
where the rotational axis coincides with the $z$-axis and the equatorial 
plane is the $x$-$y$ plane.  We assume that the configuration of a
component star is symmetric about the $x$-$y$ and $x$-$z$ planes.  
The distribution of the velocity potential is symmetric about the $x$-$y$ 
plane but anti-symmetric about the $x$-$z$ plane.  The centers of the stars 
are, therefore, on the $x$-axis.  For the equal mass IDR (NS--NS) binary 
systems, we also assume symmetry about the $y$-$z$ plane for the density and 
anti-symmetry about the $y$-$z$ plane for the velocity potential.  In actual 
computations, we introduce a spherical coordinate system 
$(r, \theta, \varphi)$ whose origin is fixed at the geometrical center of 
the gaseous component star (NS) on the $x$-axis with a distance from the 
rotational axis $d_{\ns}$ as follows: 
\begin{equation} \label{gdis}
d_{\ns} \, \equiv \, {R_{\ouou} \, + \, R_{\inin} \over 2}\ , 
\end{equation}
where $R_{\ouou}$ and $R_{\inin}$ are distances from the 
rotational axis to the inner and outer edges of the star on the $x$-axis, 
respectively.  The angle $\theta$ is the zenithal angle measured from the 
positive direction parallel to the $z$-axis and the angle $\varphi$ is 
the azimuthal angle measured from the positive direction of the $x$-axis.
Relations between the Cartesian coordinates and the spherical coordinate 
systems are expressed as follows:
\begin{eqnarray}\label{sphcor}
x\,&=&\,d_{\ns} \,+\, r\,\sin\theta \,\cos\varphi \ ,
\nonumber\\
y\,&=&\,d_{\ns} \,+\, r\,\sin\theta \,\sin\varphi \ , \\
z\,&=&\,r \, \cos \theta \, .
\nonumber
\end{eqnarray}
Since deformation of the stellar surface from a
sphere is not so large for binary stars, we may consider the stellar
surface as a single-valued function of $\theta$ and $\varphi$ and write it as
$R(\theta,\varphi)$ in this coordinate.
Since we need to solve the equation with the boundary condition 
(\ref{intbou}) {\it on the surface}, the function $R(\theta,\varphi)$ 
plays an important role in our numerical method.  
%

In this spherical coordinate system (\ref{sphcor}), we can expand the Green's 
function of the gravitational potential and the velocity potential, 
$1 / \mid {\bf r}-{\bf r}^{'} \mid$, by using the Legendre polynomials.  
For the contribution from its own component, it is expanded as 
\begin{equation} \label{legen}
{1 \over \left|\, {\bf r}-{\bf r}^{'} \,\right|}\,=\,
\sum_{n=0}^\infty f_n(r,r')\,
P_n(\cos\beta(\theta,\varphi;\theta'\varphi'))\ ,
\end{equation}
where $f_n(r,r')$ is defined as, 
\begin{eqnarray}
\!\!\!\!\!\!\!
f_n(r,r')  = 
\left\{
\begin{array}{cc}
\lefteqn{{1 \over r}\left({r' \over r}\right)^n 
\quad {\mathrm for} \quad r' \le r \ ,} \\ \\
\lefteqn{{1 \over r'}\left({r \over r'}\right)^n
\quad {\mathrm for} \quad r \le r' \ .} 
\end{array}
\right.
\end{eqnarray}
Here $P_n$ is the Legendre function and 
$\cos\beta(\theta,\varphi;\theta'\varphi')$ is defined as
\begin{equation} \label{apbeta}
\cos\beta(\theta,\varphi;\theta'\varphi')\,=\, 
\cos\theta\cos\theta'+\sin\theta\sin\theta'\cos(\varphi-\varphi')\ .
\end{equation}
Concerning the gravitational potential for the IDR systems, the contribution 
from the other component becomes as follows:
\begin{equation} \label{extgr}
{1 \over \left|\, {\bf r}-{\bf r}^{'} \,\right|}\,=\,
\sum_{n=0}^\infty {r^{'n} \over D^{n+1}}\,
P_n(\cos\gamma(\theta,\varphi;\theta'\varphi'))\ ,
\end{equation}
where $D$ and $\cos\gamma$ are defined as
\begin{equation} \label{dist}
D\,=\,\left\{(2d_{\ns})^2\,+\,r^2\,+\,2d_{\ns}\,r\sin\theta\cos\varphi
\right\}^{1/2} \ ,
\end{equation}
and
\begin{equation} \label{apgam}
\cos\gamma(\theta,\varphi;\theta'\varphi')\,=\,
{2\,d_{\ns}\,\sin\theta'\cos\varphi'\,+\, r\,
\cos\beta(\theta,\varphi;\theta'\varphi') \over D}\ ,
\end{equation}
respectively.  In this case the dashed variables should be considered
to belong to the other component star.  

Finally, several terms in the basic equations are written explicitly 
in this coordinates as follows:   
\begin{equation}\label{grad2}
\left|\,\nabla \Phi \,\right|^2\, = \,
\left({\partial\, \Phi \over\partial\,r}\right)^2
\,+\,\left({1\over r}{\partial\, \Phi \over\partial\,\theta}\right)^2
\,+\,\left({1\over r \sin\theta}
{\partial\, \Phi \over\partial\,\varphi}\right)^2\ , 
\end{equation}
\begin{equation}\label{term1}
({\bf \Omega}\times{\bf r})\cdot\nabla \Phi \, = \,
\Omega\,{\partial\, \Phi \over\partial\,\varphi}\, + \,
\Omega\, d_{\ns} \left(
\sin\theta\sin\varphi\,{\partial\, \Phi \over\partial\,r}\, + \,
{\cos\theta\sin\varphi\over r}\,{\partial\, \Phi \over\partial\,\theta}\, + \,
{\cos\varphi \over r \sin\theta}\,{\partial\, \Phi \over\partial\,\varphi}
\right) \ .
\end{equation}

\subsection{Surface fitted coordinate}

In our actual computations, we transform the basic equations into the 
surface fitted spherical coordinates (see e.g. \cite{em85} ; \cite{ue96}).  
New coordinates $(r^*,\theta^*,\varphi^*)$ are defined by 
\begin{equation}\label{ccor}
r^* = {r \over R(\theta,\varphi)}\ , \quad 
\theta^* = \theta\ , \quad {\mathrm and } \quad
\varphi^* = \varphi\ .
\end{equation}
By this transformation, the stellar interior is mapped into a 
computational domain given by 
\begin{equation} \label{cdom}
r^* \in [0,1]\ ,\quad
\theta^* \in [0,\pi]\ ,\quad {\mathrm and}\quad
\varphi^* \in [0,2\pi]\ .
\end{equation}
Accordingly, the derivatives are transformed as follows: 
\begin{mathletters}
\begin{eqnarray} \label{cderi}
{\partial \over \partial\, r} \  &\longrightarrow& \
{\cal D}_r^* \,=\,{1 \over R}\,{\partial \over \partial\,r^*}\ ,
\\&&\nonumber\\
{\partial \over \partial\, \theta} \  &\longrightarrow& \
{\cal D}_\theta^* \,=\,{\partial \over \partial\,\theta^*}\,-\,
{r^* \over R}\,{\partial R\over \partial\,\theta^*}\,{\cal D}_r^*\ ,
\\&&\nonumber\\
{\partial \over \partial\, \varphi} \  &\longrightarrow& \
{\cal D}_\varphi^* \,=\,{\partial \over \partial\,\varphi^*}\,-\,
{r^* \over R}\,{\partial R\over \partial\,\varphi^*}\,{\cal D}_r^*\ .
\end{eqnarray}
\end{mathletters}
Since transformation of the equations into this surface fitted 
coordinates (\ref{ccor}) is straightforward, we do not explicitly write 
them down in this paper.  

The disadvantage of utilizing this surface fitted coordinate is that we 
cannot decrease the number of floating calculations which appear in the 
integral equations (\ref{intpot}) and (\ref{intform}) by arranging the 
order of computations.  So it requires a large amount of CPU time and hence 
we cannot treat a large number of grid points for numerical computations.  
On the other hand, the advantage of this coordinate is that it is easy to 
treat the derivatives on the surface and also the surface itself.  
Consequently, it improves the robustness and accuracy of the computational 
scheme.  Furthermore, we do not need to treat vacuum regions.  Thus this 
advantage cancels out the disadvantage of a small number of mesh points.  

\subsection{Non-dimensional variables and choice of parameters}

The choice of the parameters in computing the equilibrium configurations 
is crucial to construct a robust numerical method.  In our actual 
computations, we use non-dimensional variables as follows:
\begin{eqnarray} \label{normq}
{\tilde {\bf r}}\,=\,{{\bf r}\over R_0}\ ,\quad
{\tilde \nabla}\,=\,R_0\,\nabla\ ,\quad
{\tilde \triangle}\,=\,R_0^2\,\triangle\ ,\quad
{\tilde \rho}\,=\,{\rho \over \rho_c}\ ,\quad
{\tilde \Theta}\,=\,{\Theta \over \rho_c^{1/N}}\ ,\nonumber\\
{\tilde p}\,=\,{p \over p_c}\ , \quad
{\tilde {\bf v}}\,=\,{{\bf v} \over \sqrt{G\rho_c}\, R_0}\ ,\quad
{\tilde \Phi}\,=\,{\Phi \over \sqrt{G\rho_c}\, R_0^2 }\ ,\quad
{\tilde \Omega}\,=\,{\Omega \over \sqrt{G\rho_c}}\ ,\\
{\tilde \phi}\,=\,{\phi \over G\rho_c R_0^2}\ ,\quad
{\tilde C}\,=\,{C \over G \rho_c R_0^2}\ ,\quad
{\tilde \beta}\,=\,{p_c \over G\rho^2 R_0^2}\ ,\quad
{\tilde a}_{lm}\,=\,{a_{lm} \over R_0^{l-2}}\ ,\nonumber
\end{eqnarray}
where $R_0$ is the geometrical radius of the gaseous component star (NS) 
along the $x$ axis, that is $R_0=(R_{\ouou}-R_{\inin})/2$, $\rho_c$ 
and $p_c$ are the density and the pressure at the coordinate center $r=0$ 
of the star.  Furthermore, an unknown quantity ${\tilde \beta}$, 
which is related to a scale of distance, appears in the Bernoulli's 
equation~(\ref{nberno}) as follows:
\begin{equation} \label{nonber}
{\tilde \Theta} \, = \, {{\tilde\beta} \over N+1}
\, \left[ \, ({\bf \tilde \Omega} \times {\bf \tilde r}) \cdot 
{\tilde \nabla} \, {\tilde \Phi} \, - \,
{1 \over 2}\,|\, {\tilde \nabla} \, {\tilde \Phi}\,|^2 \, - \, 
{\tilde \phi} \, + \, {\tilde C} \, \right] \ .
\end{equation}

For equal mass IDR (NS--NS) binary systems, in order to solve one 
stationary model, we need to specify or determine the following four 
quantities in addition to the two quantities $\Theta$ and $\Phi$:
\begin{equation}
{\tilde d}_{\ns}, \quad {\tilde \Omega}, \quad {\tilde \beta} \quad 
{\mathrm and} \quad {\tilde C}\ .  
\end{equation}
Of these quantities, we can specify the value of ${\tilde d}_{\ns}$ arbitrary.
The other three quantities can be obtained by
employing the following conditions:
\begin{equation}
{\tilde R}(\pi/2,0)\,=\,1\, ,\ {\tilde R}(\pi/2,\pi)\,=\,1\, ,
\  {\mathrm and\ } \ {\tilde \Theta}(0,\theta,\varphi)\,=\,1.0\ .  
\end{equation}
From these conditions, we can solve for the three quantities as follows:
\begin{equation}\label{para1}
{\tilde \Omega} = {{1 \over 2}\left[
\left({\partial\, {\tilde \Phi} \over \partial \varphi}
\right)_{\ouou}^2\,-\,
\left({\partial\, {\tilde \Phi} \over \partial \varphi}
\right)_{\inin}^2\right]
\,+\,{\tilde \phi}_{\ouou}\,-\,{\tilde \phi}_{\inin}
\over 
(1+{\tilde d}_{\ns})
\left({\partial\, {\tilde \Phi} \over \partial \varphi}
\right)_{\ouou}\,-\,
(1-{\tilde d}_{\ns})
\left({\partial\, {\tilde \Phi} \over \partial \varphi}
\right)_{\inin} }\ ,
\end{equation}
\begin{equation}\label{cpnorm}
{\tilde C}\,=\,-\,{\tilde \Omega}(1+{\tilde d}_{\ns})
\left({\partial\, {\tilde \Phi} \over \partial \varphi}
\right)_{\ouou}\,+\,
{1 \over 2}
\left({\partial\, {\tilde \Phi} \over \partial \varphi}
\right)_{\ouou}^2
\,+\,{\tilde \phi}_{\ouou} \ , 
\end{equation}
\begin{equation}\label{para3}
{\tilde \beta}\,=\,{N+1\over\left[ \,
 ({\bf \Omega} \times {\bf r}) \cdot \nabla \, \Phi \, - \,
{1 \over 2} \left|\, \nabla \, \Phi \,\right|^2 \, - \, \phi \, + \, C\,
 \right]_c} \ , 
\end{equation}
where subscripts {\it out}, {\it in} and $c$ mean the quantities
at the outer edge, at the inner edge and at the geometrical center of the 
star, respectively.  

For IRR (BH--NS) binary systems, since we will treat non-equal mass
configurations, in order to get one stationary model, we need to specify 
or determine the following six quantities in addition to the two quantities 
$\Theta$ and $\Phi$:
\begin{equation}
{\tilde d_{\ns}}, \quad {\tilde d_{\bh}}, \quad {\tilde \Omega}, \quad 
{\tilde \beta}, \quad {\tilde C}, \quad {\mathrm and} \quad {\tilde M_{\bh}}
\ ,
\end{equation}
where two distances from the rotational axis to each component star are
expressed as $d_{\ns}$ and $d_{\bh}$.  These distances appear in the 
gravitational potential of the point source, i.e. equation~(\ref{soupot}),
\begin{equation}
\phi_{\bh} = - {GM_{\bh} \over D_{\bh}}\ , 
\end{equation}
where
\begin{equation}
D_{\bh}\,=\,\left\{(d_{\bh}+d_{\ns})^2\,+\,
r^2\,+\,2(d_{\bh}+d_{\ns})\,r\,\sin\theta\cos\varphi\right\}^{1/2}\ .
\end{equation}

Among these six parameters, two quantities can be freely specified:
the separation between a point source and a gaseous star,
${\tilde d}_{\ns}+{\tilde d}_{\bh}$, and the mass ratio of two components,
$M_{\ns}/M_{\bh}$.  Three quantities $\tilde \Omega$, $\tilde C$ and 
$\tilde \beta$ can be determined from the same equations above.  Since 
the motion of the point mass which is subject to the gravitational force 
of the gaseous component star should be Keplerian, we can impose 
the following relation: 
\begin{equation}\label{para4}
{\tilde d}_{\bh}{\tilde \Omega}^2\,=\,\sum_{n=0}^\infty
{{\tilde I}_n \over ({\tilde d}_{\ns}+{\tilde d}_{\bh})^{n+2}}\ , 
\end{equation}
where ${\tilde I}_n$ is defined as follows, 
\begin{equation}\label{para5}
{\tilde I}_n \,=\,(n+1)\,\int_0^1 \tilde{r}^{'2}d\tilde{r}'
\int_0^{\pi}\sin\theta \,d\theta \int_0^{2\pi} d\varphi\,\,
{\tilde r}^{'n}\,{\tilde \rho}(\tilde{r}',\theta',\varphi')\,
P_n(\cos\gamma(\pi/2,\pi;\theta',\varphi'))\ .  
\end{equation}
These equations (\ref{para1}), (\ref{cpnorm}) and (\ref{para3}), and 
(\ref{para4}) for the IRR systems are simultaneously solved in the 
computational scheme.  

\subsection{Discretization and iteration procedure}

We discretize our basic equations, which are composed of 
the differential-integral equations, on the equidistantly spaced grid 
points in the surface fitted coordinates.  The derivatives are approximated 
by using the ordinary central difference scheme of the second order and the 
integrations are performed by employing the trapezoidal formula.  Let us 
denote discrete mesh points as $(r_i^*,\theta_j^*,\varphi_k^*)$ where 
$0 \le i \le N_r\, ,\  0 \le j \le N_\theta \ $ and $ \  0 \le k \le 
N_\varphi$ and $N_r$, $N_{\theta}$ and $N_{\varphi}$ are certain numbers 
employed in the approximation.  In our actual computations, we have used 
the following number of grid points: $(N_r,N_\theta,N_\varphi)=(16,8,16)$.  
The number of terms used in the Legendre expansion which appears in 
equations~(\ref{legen}), (\ref{extgr}) and (\ref{para5}) is taken into 
account up to $N_l=10$ instead of infinity.  

We apply the iteration scheme to these discretized equations.  It is 
basically identical to the self-consistent-field (SCF hereafter) method 
developed by Ostriker and Mark~(1968) and extended so as to compute 
various equilibrium configurations of self-gravitating stars by 
Hachisu~(\cite{ha86} ; \cite{keh89}).  Although there is no rigorous proof 
for the convergence to a solution and its convergence becomes slower for 
some complicated systems (e.g. \cite{ye97}),  the SCF scheme has been
proved very powerful for solving structures of rapidly rotating stars
as far as proper choice of model parameters is made.  Here we have succeeded 
in finding proper parameters and have been able to construct a computational 
code which works well for our present problem.  It should be noted that
the memory and the CPU time required for the SCF scheme are relatively small,
although we need to use not a small amount of the memory and the CPU time to 
get accurate 3D configurations.

We will briefly explain our iteration scheme.  As seen from our basic 
equations~(\ref{nberno}) and (\ref{intform}), we can symbolically 
express the discretized versions of these basic equations as follows.  
Let us denote the basic variables $\Theta$ and $\Phi$ by $Q_i$ where 
subscript $i$ means the discretized value of the quantity $Q$ at the
i-th mesh point.  The discretized basic equations are written as 
\begin{equation} \label{nrvecc}
Q_j \,=\, F_j[Q_i] \ ,
\end{equation}
where $F_j$ means the non-linear functional of variables.  These equations 
must be organized consistently. In other words, the total number of 
$Q_i$ needs to be the same as that of $F_j$.  

In our iteration scheme or, in general, in the SCF scheme, this equation
is used as follows:
\begin{equation} \label{nrvec}
Q^{(\mathrm new)}_j \,=\, F_j[Q^{(\mathrm old)}_i] \ . 
\end{equation}
Therefore, if we have some initial guesses for physical quantities,
we will be able to obtain new values for the quantities by substituting
the approximate values of physical quantities into the r.h.s of equation 
(\ref{nrvec}).  Hopefully newly obtained values for variables on l.h.s. 
can be improved ones. 

Our actual iteration procedure of the scheme is organized as follows: 
\begin{enumerate}
\item[(i)] preparation of initial guesses for $\Theta$ and $\Phi$, 
\item[(ii)] computation of values for $\rho$ and $\phi$ and gradients 
of $\Phi$ given in step(i), 
\item[(iii)] computation of new values for $\Theta$ by using equation 
($\ref{nberno}$),
\item[(iv)] computation of new values for $R$ by using values of $\Theta$, 
\item[(v)] computation of values for $\rho$ and $\phi$ and gradients of 
$\Phi$ and $\Theta$, 
\item[(vi)] computation of new values for $\Phi$ by using equation 
($\ref{intform}$), 
\item[(vii)] computation of gradients of $\Phi$ again, 
\item[(viii)] computation of new values for parameters, 
\item[(ix)] checking of convergence of all quantities; if not, go back 
to step (iii).
\end{enumerate}
As mentioned before, although there are no rigorous mathematical theorems for 
convergence of the SCF method, we have succeeded in constructing a scheme 
which accomplishes a convergence by using the above procedure.  We have used a 
certain parameter for improvement as follows: 
\begin{equation}
Q^{(\mathrm new)}_j \,=\, b\,F_j[Q^{(\mathrm old)}_i] \,+\,
(1-b)\,Q^{(\mathrm old)}_j  \ ,
\end{equation}
where the first term of r.h.s. is the improved quantities and $b$ is a 
numerical factor which is introduced to avoid jumping of solutions to
very different values and to make the iteration converge. Thus we will
call this a convergence factor.  We have set $b=1/4\sim1$ 
and used relatively smaller values for polytropes with larger $N$.  
Typically we obtain converged solutions after $20\sim 50$ iterations 
for IDR models or IRR models with nearly equal masses.  For IRR 
models with larger mass ratio, $40\sim 80$ iterations were needed.  
The CPU time was approximately 5.2 minutes per 1 iteration for 
IDR binary systems and 2.9 minutes per 1 iteration for 
IRR binary systems by using the Fujitsu VX/1R vector computer. 

\subsection{Solving method for the homogeneous term of the velocity
potential}

Finally we present the solving method for the homogeneous term of the 
velocity potential.  Since the first term alone in equation~(\ref{intform})
does not satisfy the boundary condition on the surface, we need to add
the second term, $\chi({\bf r})$, i.e. the homogeneous solution of the 
Laplace equation.  In the spherical coordinates, homogeneous solutions
to equation~(\ref{homoge}) can be expressed by using the spherical harmonic 
functions as follows:   
\begin{equation}\label{sph}
\chi(r,\theta,\varphi)\,=\,\sum_{l=1}^\infty\sum_{m=1}^l\,
a_{lm}\,r^l\,\left[1+(-1)^{l+m}\right]\,Y_l^{\ m}(\cos\theta)\,
\sin\,m\varphi\ , 
\end{equation}
where $a_{lm}$'s are certain constants and $Y_l^{\ m}(\cos\theta)$ are the 
spherical harmonic functions.  Here we have taken into account that solutions 
must be regular at the center and that the velocity potential is assumed 
symmetric about the $x$-$y$ plane and anti-symmetric about the $x$-$z$ plane.

These coefficients $a_{lm}$ are determined from the boundary 
condition~(\ref{intbou}) on the stellar surface.  After substituting the 
above expression~(\ref{sph}) into equation~(\ref{intbou}), we can 
obtain the coefficients $a_{lm}$ by applying the least square 
method to the discretized version of the boundary condition~(\ref{intbou}).  
To explain the procedure, we symbolically write the 
discretized boundary condition as follows: 
\begin{equation}\label{symbcon}
\sum_{l=1}^{N_l}\sum_{m=1}^{l} a_{lm}F_l^{\ m}(\theta_i,\varphi_j)\,-\,
H(\theta_i,\varphi_j)\,=\,0 \ ,
\end{equation}
where $F_l^{\ m}$ and $H$ are certain functions.  
To apply the least square method, we sum up the square of the l.h.s of 
the above equation at all grid points on the surface and obtain the following 
equation:
\begin{equation}\label{minene}
{\cal E}\,=\,\sum_{i,j}\left[\sum_{l=1}^{N_l}\sum_{m=1}^{l} a_{lm}
F_l^{\ m}(\theta_i,\varphi_j)\,-\,H(\theta_i,\varphi_j)\right]^2\,=\,0\ . 
\end{equation}
Coefficients $a_{lm}$ can be solved by finding solutions to the equations
which are derived by minimizing the value of ${\cal E}$ as follows: 
\begin{equation}\label{minlin}
{\delta\,{\cal E}\over\delta \,a_{lm}}\, =\,0 \ .
\end{equation}

Here we will explicitly write down some terms in equation~(\ref{intbou}). 
We need a gradient orthogonal to the stellar surface $R(\theta,\varphi)$.
It is written as 
\begin{equation}\label{bouter1}
\left.{\bf n}_S\cdot\nabla g\right|_S\, = \,{1\over\sqrt{h}}
\left({\partial\,g \over \partial\,r}\,-\,
{1\over R^2}{\partial\,R \over \partial\,\theta}
{\partial\,g \over \partial\,\theta}\,-\,
{1\over R^2\sin^2\theta}
{\partial\,R \over \partial\,\varphi}
{\partial\,g \over \partial\,\varphi}\right)_S\ , 
\end{equation}
where $g$ is a certain function and $h$ is defined as 
\begin{equation}\label{bouter2}
h(\theta,\varphi)\,=\,1\,+\,
\left({1\over R}{\partial\,R\over\partial\,\theta}\right)^2\,+\,
\left({1\over R \sin\theta}{\partial\,R\over\partial\,\varphi}\right)^2
\ .
\end{equation}
The term ${\bf n}_S\cdot({\bf \Omega}\times{\bf r}_S)$ is expressed as
\begin{equation}\label{bouter3}
{\bf n}_S\cdot({\bf \Omega}\times{\bf r}_S)\,=\,
{1\over\sqrt{h}}
\left\{-\Omega\,{\partial\,R \over \partial\,\varphi}\,+\,\Omega\,d_{\ns} 
\left(\sin\theta\sin\varphi\, - \,
{\cos\theta\sin\varphi\over R}\,{\partial\,R\over\partial\,\theta}\, - \,
{\cos\varphi \over R \sin\theta}\,{\partial\,R\over\partial\,\varphi}
\right)\right\} \ .
\end{equation}

\section{Sequences of irrotational binary}

Our new formulation for irrotational binary configurations explained
in the previous sections has worked nicely so that we can obtain stationary
sequences of binary stars for many different model parameters.
Since we assume polytropic stars in Newtonian gravity, one stationary
model of a binary system which consists of completely identical stars
can be specified only by two parameters: the polytropic index $N$
and the separation of two stars.  Thus for compressible IDR binary systems,
one stationary sequence is defined as a sequence with a fixed value
for the polytropic index $N$ but with changing the value of the separation.
On the other hand, for compressible IRR binary systems, we have one more 
parameter to characterize a sequence, which is the mass ratio $\ratmas$ 
of a gaseous star (NS) to a point source (BH).  

A sequence with fixed masses for component stars and with the 
prescribed equation of state can be regarded as an evolutionary sequence 
due to emission of GW since the irrotationality assures the conservation of 
circulation as discussed in section 2.1 (\cite{mi74}; \cite{efh90}).  
Concerning this evolutionary or stationary sequence, LRS1 showed that 
dynamical instability will set in at the point on the sequence where the 
total angular momentum $J(d_G)$ or the total energy $E(d_G)$ of the binary 
system attains its local minimum.  Here $d_G$ is the separation of two 
centers of mass of component stars.  They also proved that the distance 
at the minimum point for $J(d_G)$ and that for $E(d_G)$ coincide for a 
stationary sequence.  
This property of a sequence can also be applied for the present results.  
We will call this point the dynamical instability limit and denote the 
corresponding separation by $r_d$.  This dynamical instability limit may or 
may not be reached before the point of termination of the sequence of 
equilibrium models.  The minimum separation for equilibrium configuration 
is referred to as the Roche limit $r_R$.  For the equal mass IDR binary 
systems, there is another type of critical point which is the contact limit 
where the two gaseous components come to touch each other at the inner 
edges.  We denote this contact separation by $r_c$.  We will refer to these 
characteristic separations as the critical separations.  

It is important to know behaviors of $J(d_G)$ and/or $E(d_G)$ along stationary 
sequences because final states of evolution will be affected by the relation 
among those critical separations.  Since GW carry away
the angular momentum as well as the energy, the separation of a binary system
will decrease as the system evolves.  If $r_d$ is larger than two other 
critical separations, i.e. $r_R$ and $r_c$, two component stars will begin to 
merge on a dynamical time scale because of the orbital instability.  On the 
other hand, if $r_d$ is smaller than the other critical separations, other 
hydrodynamical phenomena is expected to occur.  In particular, for the IDR 
binary systems, the contact dumbbell configurations may appear similarly
to the {\it incompressible} case (\cite{eh85}) and for the IRR binary systems 
the Roche lobe overflow may occur.  We will discuss this matter later by 
comparing our results with those of recent hydrodynamical simulations of 
binary systems by several authors.  

In the following subsections, we will tabulate physical quantities which 
characterize the sequences.  In Tables we show the following quantities.
They are the normalized values for the separation of the geometrical centers 
of two component stars, i.e.  $\tilde{d} = 2\tilde{d}_{\ns}$ for the IDR 
systems, and $\tilde{d}=\tilde{d}_{\bh}+\tilde{d}_{\ns}$ for the IRR 
systems, the separation of mass centers of two component stars $\bar{d}_G$, 
the angular velocity $\bar{\Omega}$, the total angular momentum $\bar{J}$ 
and total energy $\bar{E}$.  These quantities are normalized as follows:
%
\begin{eqnarray} \label{pqnorm}
\bar{d}_G\,=\,{d_G \over R_N} \, \quad
{\mathrm for \ IDR\, ,}\qquad 
\bar{d}_G\,=\,{d_G \over R_N}\left({M_{\ns} \over M_{\bh}}\right)
^{1\over 3}\,\quad
{\mathrm for \ IRR\, ,} \nonumber
\\
\bar{\Omega}\,=\,{\Omega \over (\pi G\bar\rho_N)^{1/2}}\ ,\quad
\bar{J}\,=\,{J \over (GM^3R_N)^{1/2}}\, \quad{\mathrm and} \quad
\bar{E}\,=\,{T\,+\,W\,+\,U \over GM^2/R_N}\ ,
\end{eqnarray}
where $M$ is the mass of primary star, $R_N$ is the radius of the spherical 
polytrope with the same mass $M$ and with the same polytropic index $N$.  
The quantity $\bar{\rho}_N$ is defined as $\bar\rho_N\,=\,M/(4\pi R^3_N/3)$.  
These normalizations are the same as those adopted in LRS's papers.  

The quantities $T$, $W$ and $U$ are the total kinetic energy, the total 
potential energy and the total thermal energy defined as usual 
(see e.g. Tassoul 1978).  The separation between the mass centers of two 
component stars $d_G$ is computed as follows: 
\begin{equation} \label{pqdis}
d_G \,=\, 2 {\int_V \,x\,\rho \,d^3 {\bf r}^{'}\over M}
\, \quad {\mathrm for \ IDR,}\qquad
d_G \,=\,  {\int_V \,x\,\rho \,d^3 {\bf r}^{'}\over M} + d_{\bh}
\, \quad {\mathrm for \ IRR.}
\end{equation} 
The total angular momentum $J$ is computed by using the velocity potential 
as follows: 
\begin{eqnarray}
J& =&2\,\int_V \,\rho\,({\bf r}\times{\bf v})\cdot{\bf e}_z
d^3 {\bf r}^{'} \nonumber \\
&=&2\,\int_V\,\rho\,
\left\{
{\partial\,\Phi\over\partial\,\varphi}\, + \,
d_{\ns} \left(
\sin\theta\sin\varphi{\partial\,\Phi\over\partial\,r}\, + \,
{\cos\theta\sin\varphi\over r}{\partial\,\Phi\over\partial\,\theta}\, + \,
{\cos\varphi \over r \sin\theta}{\partial\,\Phi\over\partial\,\varphi}
\right)\right\}
d^3 {\bf r}^{'} \quad  {\mathrm for \ IDR,} \\
J&=&\int_V \,\rho\,({\bf r}\times{\bf v})\cdot{\bf e}_z
d^3 {\bf r}^{'}\,+\,M_{\bh}\,d^2_{\bh}\,\Omega \quad 
{\mathrm for \ IRR.} 
\end{eqnarray}
The quantity $T/\left| W \right|$ is also tabulated but note that the 
definition is different from LRS1.  The value of the virial constant 
$\VC$ normalized by the total gravitational potential is computed as 
follows:
\begin{equation}\label{virial}
\VC\, = \,{\left|\,2\,T\,+\,W\,+\,3\,U/N\,\right|\over 
\left|\, W \,\right|} \ , 
\end{equation}
where the value of $U/N$ vanishes for the models with $N=0$.  
We also show the values of $\bar{R}$ which corresponds to average radius 
defined as follows:
\begin{equation}\label{rnorm}
\bar{R}\, = \,\left({3\,V\over 4\,\pi\,R^3_N}\right)^{1\over 3}\,=\,
\left[{1\over4\,\pi\,R^3_N}\int_0^{\pi}\sin\theta\,d\theta \int_0^{2\pi} 
d\varphi\,R^3(\theta,\varphi)\right]^{1\over 3}\ . 
\end{equation}
Generally, this quantity increases as the separation decreases.  It 
indicates that the volume of the gaseous star always increases and, 
accordingly, the central density decreases as the two component star 
approaches during the evolution.  

\subsection{Compressible IDR binary systems with equal mass}  

\subsubsection{Accuracy of the new numerical method and stationary 
sequences}  

We have computed several equal mass IDR (NS--NS) sequences.  In Tables 
\ref{irdrn0.tab}-\ref{irdrn15.tab}, the physical quantities are shown for 
sequences with polytropic indices $N=0,\ 0.5,\ 0.7,\ 1\ {\mathrm and}\ 1.5$,
respectively.  
\placetable{irdrn0.tab}
\placetable{irdrn05.tab}
\placetable{irdrn07.tab}
\placetable{irdrn1.tab}
\placetable{irdrn15.tab}

In order to check on the accuracy of our new numerical method, we compare 
our results with those of LRS2 who used the ellipsoidal approximation which 
should be excellent at least when the separation is relatively large.  
In Figure~\ref{irdrn15.gra}, $\bar{J}(\bar{d}_G)$, $\bar{E}(\bar{d}_G)$ 
and $\Oseq$ are plotted for the $N=1.5$ sequence.  Similar figures for 
the smaller values of polytropic indices $N=0,\ 0.5\ {\mathrm and}\ 1$ are 
presented in Paper I.  In this figure our results for the $N=1.5$ 
sequence agree well with those in LRS2 within a relative error less than 
$0.8 \%$  for $\Jseq$ and $0.4 \%$ for $\Eseq$.  This agreement with the 
LRS's results shows that our method can accurately compute configurations 
with any polytropic index as far as deformation is not far from ellipsoid.  
In Figure~\ref{irdrall.gra}, we draw $\Jseq$ and $\Oseq$ for sequences with 
$N=0,\ 0.5,\ 0.7,\ 1\ {\mathrm and}\ 1.5$.  We also plot the results of 
incompressible ($N=0$) models computed by using a different formulation and 
a different solving method described in Paper II.  The results of $N=0$ 
sequences computed by the present method and those obtained in paper II 
coincide very well, although these two results are computed by using two 
totally independent codes.  This fact shows that our present computational 
code works accurately even for highly deformed configurations.  In particular,
since our new code is accurate enough to resolve curves around the turning 
point of $\Jseq$ or $\Eseq$ for $N=0$ polytropes, we can expect that it will 
give precise behavior around the critical separations for $N\ne 0$ polytropes
as well.  As discussed in Paper II, the point of the minimum value of $\Jseq$ 
or $\Eseq$ does not appear along the binary sequence when the polytropic 
index satisfies the condition $N\ga0.7$.  Since the equation of state of 
realistic neutron stars is thought to be approximated by polytropic 
equation of state with index $0.5\sim 1$, there arises a possibility that
in a new scenario of the inspiraling neutron star binary in weak gravity 
limit, two component stars approach each other and finally form a dumbbell 
like configuration before dynamical instability sets in.  
\placefigure{irdrn15.gra}
\placefigure{irdrall.gra}

As mentioned above and in Paper I, we have shown that dynamical instability 
limit disappears along binary sequences for polytropes with $ N \ga 0.7$.  We 
tabulate values of $r_d$ and $r_c$ of our results and those of LRS2 for 
various polytropic indices in Table~\ref{ISCO.tab}.  Quantitative 
difference becomes larger as the polytropic index $N$ becomes larger.  
Even for smaller values as $N\sim0.5$, values of the angular 
velocity $\bar{\Omega}(r_d)$ differ by $\sim 10 \%$ or so between our present 
results and the results of LRS2.  The difference of quantities at $r_c$ 
is much larger.  We will explain the reason why the ellipsoidal 
approximation cannot give precise values for $r_d$ or $r_c$ in below.  
\placetable{ISCO.tab}

\subsubsection{Binary configurations}

In Figures~\ref{irdrn05.con}, \ref{irdrn1.con} and \ref{irdrn15.con}, we show 
contours of the density and the velocity potential distributions for $N=0.5$, 
$N=1$ and $N=1.5$ polytropes in contact phases, respectively.  It is noted 
that the configuration with $N=0.5$ in Figure~\ref{irdrn05.con} is 
dynamically unstable.  
\placefigure{irdrn05.con}
\placefigure{irdrn1.con}
\placefigure{irdrn15.con}

We can clearly see differences between the present results and those
of the ellipsoidal approximation from these figures.  In general, the 
density distribution of a component star with a larger polytropic index 
is centrally condensed, i.e. the star with a soft equation of state consists 
of a high density core and an extended low density envelope.  Although 
the central high density cores can be approximated well by the ellipsoidal
configurations, the extended envelopes of compressible gaseous stars deform
from ellipsoidal figures by tidal force of the other component.  By comparing 
panels~(a) and (b) in Figures~\ref{irdrn05.con}, \ref{irdrn1.con} and 
\ref{irdrn15.con}, the location of the maximum density shifts to a larger 
value of $x$.  It implies that the deformation of the envelope due to the 
tidal force becomes large and that the inner parts of the envelopes come 
to contact each other.  Because of this significant deformation of the 
envelopes, two soft components stars contact each other at a larger 
separation $\bar{d}_G=r_c$ than that for the ellipsoidal approximation.  
Therefore differences between our results and those of LRS2 for $r_c$ or 
$r_d$ become significant.  

The good agreement of our results and those of the ellipsoidal approximation
as seen from Figure~\ref{irdrn15.gra} can be explained by using these 
density distributions as follows.  First, since the mass content in the 
envelope is very small, the contribution of the envelope to those
quantities is very small.  Furthermore, larger values of $r_c$ means that
the condensed core region becomes less affected from tidal force and 
consequently the shape deforms little from an ellipsoidal figure.  Therefore 
those quantities which are determined mainly from the contribution of the core 
region coincide well with the results of the ellipsoidal approximation.  
It should be noted, however, that precise values of $r_c$ or $r_d$ cannot 
be computed by the ellipsoidal approximation as mentioned above.  

From panels~(c) of these contours with several polytropic indices, 
we can see that the cross section parallel to the $y$--$z$ plane is slightly 
{\it prolate}, i.e. the radius along the axis ${\tilde Z}$ is longer than 
that along the ${\tilde Y}$ axis.  This difference of the radii is smaller 
for the star with larger polytropic indices.  The cross section becomes 
almost axisymmetric about the $x$--axis for polytropes with $N=1$ and 
$N=1.5$.  Concerning the velocity fields, we show distributions of the 
velocity potential in the figure~(d) of each Figure.  Since the contour of 
the velocity potential seems almost paralell to $z$-axis i.e. 
$z$-independent,  the velocity field is almost plane parallel to equatorial 
plane and do not depend on $z$ also.  

In Figure~\ref{irdrn05.con2}, we show the internal structure for $N=0.5$ 
and $\tilde{d}=2.4$ which almost corresponds to the configuration at 
the dinamical instability limit $r_d$.  Different from Figures~
\ref{irdrn05.con}, \ref{irdrn1.con} and \ref{irdrn15.con}, the shape 
of contours in panels (a), (b) and (d) does not have a cusp, since two 
component stars are slightly detached.  In Figure~\ref{irdrn05.con2}~(e) 
contours of the velocity potential in the plane which is parallel to 
the $y$-$z$ plane and intersects with the geometrical center of the star 
are shown.  As mentioned before, the velocity potential almost do 
not depend on $z$.  Therefore, we can consider that this figure expresses 
that the flow is almost planar, i.e. parallel to the equatorial plane, and 
independent of $z$ cooridnate.  The stationary velocity field {\bf u} in 
the equatorial plane in the rotational frame is also shown in 
Figure~\ref{irdrn05.con2}~(f).  The shapes of the star whose internal 
structures are shown in Figure~\ref{irdrn05.con2} are displayed in 
Figure~\ref{irdrn05.surf}.
\placefigure{irdrn05.con2}
\placefigure{irdrn05.surf}

In Figure~\ref{irdrvec.prof}~(a) we show the velocity profile along the 
$x$-axis for the polytropes with $N=0,\ 0.5,\ 1$ and $1.5$ at contact phases, 
and also that for $N=0.5$ polytropes at nearly the dynamical instability limit 
in Figure~\ref{irdrvec.prof}~(b).  Along the $x$-axis, there exists only 
the $y$-component of the velocity field $v_y$ in the Cartesian coordinate.  
For synchronously rotating stars, $v_y$ is proportional to $x$, i.e.
$v_y \propto x$, on the $x$-axis.  On the other hand, $v_y$ tends to a 
constant value at larger $x$ for the irrotational binary systems.  If we 
approximate the distribution along the $x$-axis by 
$v_y \propto x^\alpha + const$, the exponent $\alpha$ is $\alpha \sim 0.4$ 
at the inner region and $\alpha \sim 0.1$ at the outer region near the 
contact stage.  The significant difference from synchronously rotating 
binaries is that the velocity $v_y$ at $x=0$ for the contact phase does not 
vanish but is finite.  These differences about the velocity fields will 
result in the different evolution from that of synchronously rotating 
binaries after the contact or merging phase of two stars.  
\placefigure{irdrvec.prof}

\subsection{Compressible IRR binary systems with several mass ratios}  

\subsubsection{Stationary sequences}  

We have computed IRR (BH--NS) sequences for several polytropic indices 
and several mass ratios.  In Tables~\ref{irrreqmn0.tab}-\ref{irrrm10n1.tab},
the physical quantities are shown for polytropic sequences with $N=0,\ 0.5\ 
{\mathrm and}\ 1$ for $\ratmas=1,\ 0.5$ and $0.1$ binaries, respectively.  
\placetable{irrreqmn0.tab}
\placetable{irrreqmn05.tab}
\placetable{irrreqmn1.tab}
\placetable{irrrm2n0.tab}
\placetable{irrrm2n05.tab}
\placetable{irrrm2n1.tab}
\placetable{irrrm10n0.tab}
\placetable{irrrm10n05.tab}
\placetable{irrrm10n1.tab}

In Figures~\ref{irrreqmn0.gra} and \ref{irrrm10n0.gra}, we show three 
different results for $ N = 0$ polytropes: the results computed by our 
present method, those of LRS1 and those obtained in Paper II.  For equal 
mass sequences in Figure~\ref{irrreqmn0.gra}, our present results 
agree very well with those of Paper II everywhere along the sequence down 
to the smallest separation.  Therefore, just as the IDR binary systems, 
our computational method will be able to give accurate models for the 
contact binary systems with highly deformed stars around the critical 
radius.  In Figure~\ref{irrrm10n0.gra}, we can see a good agreement
of our present results with other two results for the sequence with
$\ratmas = 0.1$.  Since the mass of a point source $M_{\bh}$ is 
dominant, deformation of a gaseous component (NS) does not affect the 
physical quantities shown in those figures.  Therefore three different 
results agree well even near the critical separations.  However the 
critical separations $r_d$ and $r_R$ of our present results agree with 
those of Paper II but not with those of LRS1.  Therefore the ellipsoidal 
approximation gives different values for the critical separations again
as is expected.  
\placefigure{irrreqmn0.gra}
\placefigure{irrrm10n0.gra}

In Figures~\ref{irrreqmn1.gra} and  \ref{irrrm10n1.gra}, we draw the 
sequences for $N = 1$ polytropes with $\ratmas = 1$ and $0.1$, respectively.  
In these figures, results by LRS1 are also shown.  As seen from these 
figures, for polytropes with larger $N$ and/or larger $\ratmas$, two results 
agree each other everywhere.  Even for smaller values of $N$ and 
$M_{\ns}/M_{\bh}$, these results are in good agreement for models with 
larger separations.  Hence our code for the IRR binary systems works 
accurately as that for the IDR binary systems as mentioned before.  In 
these figures, however, there are clear differences from those for $N=0$ 
polytropes.  In our {\it present} calculations, the dynamical instability 
limit does not appear along the sequence with $N=1$, i.e. both $\Jseq$ and 
$\Eseq$ curves terminate at the smallest separation without reaching a 
turning point.  
\placefigure{irrreqmn1.gra}
\placefigure{irrrm10n1.gra}

However we should note that situation for the IRR binary systems is 
different from that for the equal mass IDR binary systems.  For the equal 
mass IDR binary systems, the end point of the sequence exactly corresponds 
to a configuration with $\tilde{d}=2$.  Thus we know $r_c$ a priori.  
However, for the IRR binary systems, there exist sequences for which it is 
hard to find turning points along stationary sequences.  This may occur for 
models with larger mass difference, say $\ratmas\sim 0.1$, because they 
have `shallow' local minimum as in Figure~\ref{irrrm10n0.gra}~(a) or (b).  
Consequently we need to compute the terminal point of the sequence by 
changing the separation $\tilde{d}$ carefully.  

In Figure~\ref{irrrm2all.gra}, we show the IRR sequences of polytropes
with the same mass ratio $\ratmas=0.5$ but with different polytropic indices 
$N=0,\ 0.1,\ 0.3,\ 0.5$ and $1$.  The critical points of the sequences can be 
found from these figures.  Since the curves for smaller values of $N$ show 
clearly the turning points of $\Jseq$ in Figure~\ref{irrrm2all.gra}~(a),
the sequences around the smallest separation have been considered reliable.  
We also show the results of Paper II by dots as before which also show the 
accuracy of our present computations.  Figure~\ref{irrrm2all.gra}~(a) 
reveals that the turning points disappear for 
$N \ga 0.6$ polytropes  
and hence the sequence directly reaches the Roche-Riemann limit with $r_R$ 
as the BH and the NS approach each other.  The realistic equation of state 
is considered to be included in this range of the polytropic index.  These 
results suggest a new scenario for the final fate of the IRR binary system 
evolution due to emission of GW.  In the limit of Newtonian gravity, the 
IRR binary systems will not become dynamically unstable before reaching 
the Roche-Riemann limit with $r_R$.  Therefore the NS will not `fall on' to 
the BH but the tidal disruption or Roche overflow will happen.  Systematic 
studies on the Roche limit $r_R$ and ISCO of the BH--NS system in weak 
gravitational limit with various mass ratios should be considered in the 
subsequent paper (\cite{ue98c}).
\placefigure{irrrm2all.gra}

\subsubsection{Binary configurations}

The compressible gaseous star components (NS) become highly deformed for 
larger mass ratios.  In Figures~\ref{irrrm2n05.con} and \ref{irrrm2n1.con},
contours of the density and the velocity potential distributions for binary 
systems with $\ratmas = 0.5$ are displayed for polytropic components with 
$N=0.5$ and $1$, respectively.  They also correspond to the models with 
the smallest separations tabulated in Tables~\ref{irrrm2n05.tab} and 
\ref{irrrm2n1.tab}, respectively.  In panels~(a) and (b) of 
Figures~\ref{irrrm2n05.con} and \ref{irrrm2n1.con}, the density 
distributions in the equatorial and meridional sections show that the inner 
edges of the stars become cusp-like.   It should be noted that these 
configurations are those of near the smallest separations and of dynamically 
stable models.  They also show that the configurations become slightly 
prolate for the IRR binary systems.  The dependency of physical quantities 
on the values of the polytropic index $N$ is the same as that of the IDR 
systems.  Consequently the curves of $\Jseq$ and $\Eseq$ agree well with 
those of LRS but the critical distances $r_d$ and $r_R$ become different 
from theirs.  For the smaller $\ratmas$, the difference from the results 
of LRS becomes relatively smaller because the point source with a larger 
mass determines the bulk structures and the quantities such as $\Jseq$ and 
$\Eseq$.  
\placefigure{irrrm2n05.con}
\placefigure{irrrm2n1.con}

In Figure~\ref{irrrm2n05.con}(e), the distribution of the velocity potential 
in the $\varphi=\pi/16$ and $\varphi=17\pi/16$ planes is shown.  We can see 
there is almost no $z$-dependence of $\Phi$ in these planes also as 
mentioned in the previous section.  
In Figure~\ref{irrrm2n05.con}~(f), the velocity field in the rotating frame 
$\bf{u}$ of the same model is drawn.  Since the density of the gas becomes 
relatively low near the inner edge of the star, the velocity becomes slower 
there accordingly.  In Figure~\ref{irrrm10n05.surf}, we show the shape of 
the surface of the model with  $N=0.5$ and $\ratmas=0.1$ at the smallest 
separation displayed in Table~\ref{irrrm10n05.tab}.
\placefigure{irrrm10n05.surf}

\section{Discussion}

\subsection{Equilibrium approach vs. dynamical simulation}

In this paper we have presented a new numerical method based on a new 
formulation to compute the irrotational gaseous binary systems and showed 
results for BH--NS and NS--NS binary systems in Newtonian gravity. 

There have been two different approaches to investigate realistic close 
binary systems which are consist of compact objects : one is an equilibrium 
approach as presented in this paper and the other is a dynamical approach 
such as simulations.  {\it Synchronously} rotating NS--NS binary systems 
have been studied by several authors by using equilibrium approaches.  In 
particular, the ISCO have been investigated under the post-Newtonian 
approximation~(\cite{sh97}) and by employing simplified general relativistic 
formulation~(\cite{ba97c}).  However, for irrotational binary systems, 
only approximate solutions have been investigated in the equilibrium 
approach (LRS2).  Therefore our present method based on a new formulation 
is the first attempt to treat stationary states of irrotational 
binary systems exactly, even though Newtonian gravity is used.  

On the other hand, several researchers have performed dynamical simulations
for this problem.  This approach has been employed mainly from requirement 
of knowing wave forms of GW during dynamical coalescence 
because they are crucial to extract information about the sources by 
comparing theoretical predictions with observational results.  
In this approach, some authors have computed evolution of synchronously 
rotating coalescing binary NS systems~(\cite{sno92}; \cite{son97}; 
\cite{rs94} ; see also \cite{rs96}).  In particular, Rasio \& Shapiro~(1994) 
and Shibata, Oohara \& Nakamura~(1997) constructed initial equilibrium 
configurations of synchronously rotating binary systems accurately.  
By starting from such initial states, they carried out dynamical evolutionary
computations by using their hydrodynamical codes and compared their results 
with other results which have been obtained by equilibrium approaches.  
Shibata, Oohara \& Nakamura~(1997) have proceeded their computations
up to the first post-Newtonian (1PN) hydrodynamical simulations starting 
from equilibrium configurations of synchronously rotating binary systems 
obtained in the 1PN approximation.  They used the equilibrium 
configurations which have slightly larger and smaller separations than $r_d$ 
as their initial conditions and checked whether those configurations are
dynamically stable or not.  

However, for non-synchronously rotating systems, computations using a 
dynamical approach are not yet completely satisfactory because of very 
approximate initial configurations or having computations which are not as 
accurate as for the synchronous models.  (see for example, \cite{sno92} ; 
\cite{zcm94}, 1996; \cite{da94}; \cite{rrj97}).  Shibata, Nakamura \& 
Oohara~(1992) tried to use two axisymmetrically rotating  polytropes as 
their initial configurations to approximate irrotationally spinning binary 
configurations.  As mentioned in the previous section, since the component 
stars of irrotational binary systems become slightly prolate, such 
axisymmetric configurations are unsatisfactory.  Zhuge, Centrella \& 
McMillan~(1994) and Ruffert, Rampp \& Janka~(1997) have computed coalescence 
of non-synchronous binary systems by using approximate irrotational 
configurations as their initial conditions.  However it should be noted that 
their results have not been carefully checked yet.  Since the results of 
hydrodynamical simulations are very sensitive to initial configurations 
and/or to numerical schemes, it is necessary to check whether those 
simulations are as accurate as those of Rasio \& Shapiro~(1994) or 
Shibata, Oohara \& Nakamura~(1997).   Therefore our results of equilibrium 
approach will be important not only for checking the results of 
hydrodynamical computations but also for providing reliable initial 
conditions for irrotational binary systems.  Concerning the IRR 
configurations, the dynamical simulation is recently performed by Lee 
\& Klu\'zniak~(1997).  However, detailed comparison of our results with 
theirs will be discussed in the subsequent paper (\cite{ue98c}).

\subsection{Final stage of evolution for NS--NS or BH--NS binary systems}

As discussed before, critical distances play an essential role in the final
stage of binary evolution.  In Figure~\ref{figsummary}, we summarize our 
results and those of LRS1 for compressible irrotational binary systems
schematically.  The upper part corresponds to the IDR binary systems with
equal mass and the lower part to the IRR binary systems with 
$\ratmas = 0.5$.  In this figure arrows show the separations of two stars 
for each sequence.  Since the separation becomes smaller due to GW emission, 
the binary stars evolve from the right to the left on each arrow, i.e. from 
a distant position to a position with a small distance.  It should be noted 
that, for relatively soft equations of state, the radius at which dynamical
instability sets in does not appear along the sequence from our numerical 
results.  In particular, this is clear contrast to the results of LRS 
who always found dynamically unstable limits for the IRR binary systems.
\placefigure{figsummary}

For the equal mass IDR binary systems, several numerical simulations
show the formation of dumbbell-like configurations (\cite{sno92} ; 
\cite{zcm94} ; \cite{da94} ; \cite{rrj97}).  This can be explained by using 
the critical distances as follows.  Since the critical distance for the 
dynamical instability $r_d$ does not appear on the equal mass NS--NS binary 
sequence for softer equations of state, the system evolves to the state
with the critical radius for the contact phase, i.e. $r_c$.   This can be
also possible for the ellipsoidal models.  However, the value of $N$ is
much smaller for our exact numerical computations than that obtained from the
ellipsoidal approximation.

It may be worth while considering evolution which will probably occur 
after contact of two stars.  There arises a difference from that for 
the synchronously rotating binary systems.  Since irrotational stars have 
non-zero velocity at the inner edge in the rotational frame at the
contact stage as shown in Figure~\ref{irdrn05.con2}~(f),  the eddy 
may be excited and gases result in turbulence.  Consequently complicated
hydrodynamical phenomena are expected to happen.  Thus it is necessary to 
perform dynamical simulations with high resolution in these regions (see e.g. 
\cite{rs96}).  Some detailed discussions of the effects due to realistic 
micro physics of the coalescing NS--NS binary systems are given in 
\cite{rjts97}.  

It is also important to investigate more realistic irrotational NS-NS in 
the framework of general relativity (GR).  Shibata~(1997) 
has reported the GR effect on the synchronous NS--NS 
binary systems.  He obtained numerically exact configurations in 
post-Newtonian (PN) gravity.  According to his results, the  effect 
weakens the tidal effect and hence the turning point of $J(d_G)$ or 
$E(d_G)$ disappears along the binary sequence with the constant mass.  
Therefore it may be probable also for irrotational binary systems to reach
a contact phase without suffering from the dynamical instability when 
the GR effect is taken into account.  

One of recent topics about merging of NS-NS binary systems is the numerical 
results obtained by Wilson, Mathews \& Marronetti~(1996).  They have involved 
a simplified GR effect and concluded that individual neutron star in the binary
system would collapse to form a BH on the circular orbit before coalescence.  
Although this result has been criticized by many authors, they recently 
refute that such phenomena can occur owing to the coupled effects of the spin 
of NS and the GR effect higher than 2PN (see \cite{mmw97} and references 
there in).  A definite answer to this conjecture may be given by combining 
our method of irrotational binary systems and the GR treatment of gravity 
adopted in Baumgarte et al.~(1997c) for synchronously rotating systems 
by using the formulation of GR generalized Bernoulli 
equation (\cite{taub59} ; \cite{bgm97} ; {\cite{asa98} ; {\cite{sh98} ; 
{\cite{teu98})

Concerning the IRR binary systems, situation is somewhat different
from the IDR binary systems.  As described above, the ellipsoidal 
approximation shows that the critical radius $r_d$ always appears along 
the binary sequence, i.e. $r_R < r_d$ for any $N$.  However, from our 
results, no dynamical instability limit along the sequence appears for 
softer equations of state with 
$N \ga 0.6$.  
This implies that neutron stars will not `fall on' to the BH but that 
the Roche lobe overflow may occur. 
Further discussion for IRR binary systems will be presented in a subsequent 
paper (\cite{ue98c}).  

To investigate more realistic final fates of BH-NS binary systems, 
we must further consider the radius $r_{GR}$ at which GR instability sets 
in, i.e. the limit inside of which no stable orbits exist 
due to the effect of strong gravity.  Consequently, binary stars also fall 
down onto each other when two stars come within 
$r_{GR}\sim 6 GM_{\mathrm tot}/c^2$.  
Although this critical distance $r_{GR}$ has not been treated in our 
present paper,  we will be able to implement the GR 
effect concerning the ISCO at $r_{GR}$ in our treatment of the 
Roche--Riemann type binary systems by using the pseudo-Newtonian potential 
as was done by Taniguchi \& Nakamura~(1996).  In such a situation, 
the final fate of the BH--NS system will be determined by the positions of 
two lengths $r_R$ and $r_{GR}$ along the binary sequence.  If we will be
able to include the GR effect, we will find the position
of the radius $r_{GR}$ on this diagram~\ref{figsummary} and clarify 
the final fate of the BH--NS systems.

\acknowledgments

We would like to thank Prof. J. C. Miller for carefully reading the 
manuscript and for helpful comments.  One of us (KU) also would like to 
thank Prof. D. W. Sciama and Dr. A. Lanza for their warm hospitality at 
ICTP and SISSA.  Numerical computations were carried out by at the 
Astronomical Data Analysis Center of the National Astronomical Observatory, 
Japan.  

\clearpage

%
%
%
\begin{table}
\begin{center}
\begin{tabular}{cccccccccccccc}
\tableline
\tableline
\\[-3truemm]
$\tilde{d}$ & $\bar{d}_G$ & $\bar{\Omega}$ & $\bar{J}$ &
$\bar{E}$ & $T/|W|$ & $\VC$ & $\bar{R}$ \\
\\[-4truemm]
\tableline
\\[-4truemm]
4.0 & 4.197 & 1.914(-1) & 1.454 & -1.316 & 8.347(-2) & 6.777(-3) & 9.994(-1) \\
3.8 & 4.022 & 2.043(-1) & 1.425 & -1.321 & 8.672(-2) & 6.788(-3) & 9.994(-1) \\
3.6 & 3.852 & 2.183(-1) & 1.397 & -1.326 & 9.019(-2) & 6.828(-3) & 9.994(-1) \\
3.4 & 3.690 & 2.335(-1) & 1.371 & -1.331 & 9.388(-2) & 6.912(-3) & 9.994(-1) \\
3.2 & 3.537 & 2.495(-1) & 1.348 & -1.336 & 9.782(-2) & 6.960(-3) & 9.994(-1) \\
3.0 & 3.399 & 2.662(-1) & 1.329 & -1.340 & 1.021(-1) & 7.039(-3) & 9.994(-1) \\
2.8 & 3.279 & 2.830(-1) & 1.318 & -1.343 & 1.068(-1) & 7.218(-3) & 9.994(-1) \\
2.6 & 3.184 & 2.985(-1) & 1.315 & -1.344 & 1.117(-1) & 7.636(-3) & 9.994(-1) \\
2.4 & 3.124 & 3.110(-1) & 1.326 & -1.341 & 1.168(-1) & 8.117(-3) & 9.994(-1) \\
2.2 & 3.107 & 3.181(-1) & 1.350 & -1.334 & 1.215(-1) & 8.670(-3) & 9.994(-1) \\
2.0 & 3.131 & 3.169(-1) & 1.367 & -1.330 & 1.227(-1) & 9.209(-3) & 9.994(-1) \\
\\[-4truemm]
\tableline
\end{tabular}
\end{center}
\tablenum{1}
\caption{Stationary sequence of the equal mass IDR(NS--NS) system with $N=0$. 
$\tilde{d}$ is the separation of the geometrical centers of two gaseous 
component stars (NS) which is defined as $\tilde{d} = 2d_{\ns}/R_0$, where 
$R_0$ is the geometrical radius of the star along the $x$ axis, 
that is $R_0=(R_{\ouou}-R_{\inin})/2$ (see equations (\ref{gdis}) and 
(\ref{normq})).  Other quantities are defined in equations (\ref{pqnorm}), 
(\ref{virial}) and (\ref{rnorm})}
\label{irdrn0.tab}
\end{table}
\begin{table}
\begin{center}
\begin{tabular}{cccccccccccccc}
\tableline
\tableline
\\[-3truemm]
$\tilde{d}$ & $\bar{d}_G$ & $\bar{\Omega}$ & $\bar{J}$ &
$\bar{E}$ & $T/|W|$ & $\VC$ & $\bar{R}$ \\
\\[-4truemm]
\tableline
\\[-4truemm]
4.0 & 4.165 & 1.933(-1) & 1.446 & -1.231 & 7.618(-2) & 2.774(-2) & 1.004 \\
3.8 & 3.983 & 2.069(-1) & 1.415 & -1.236 & 7.925(-2) & 2.751(-2) & 1.004 \\
3.6 & 3.804 & 2.219(-1) & 1.385 & -1.241 & 8.255(-2) & 2.726(-2) & 1.004 \\
3.4 & 3.631 & 2.383(-1) & 1.355 & -1.247 & 8.608(-2) & 2.699(-2) & 1.004 \\
3.2 & 3.466 & 2.560(-1) & 1.327 & -1.253 & 8.986(-2) & 2.671(-2) & 1.005 \\
3.0 & 3.312 & 2.749(-1) & 1.301 & -1.259 & 9.385(-2) & 2.638(-2) & 1.005 \\
2.8 & 3.172 & 2.944(-1) & 1.280 & -1.264 & 9.806(-2) & 2.605(-2) & 1.005 \\
2.6 & 3.054 & 3.135(-1) & 1.265 & -1.268 & 1.025(-1) & 2.581(-2) & 1.006 \\
2.5 & 3.005 & 3.223(-1) & 1.260 & -1.269 & 1.047(-1) & 2.571(-2) & 1.007 \\
2.4 & 2.965 & 3.301(-1) & 1.258 & -1.270 & 1.068(-1) & 2.558(-2) & 1.008 \\
2.3 & 2.935 & 3.366(-1) & 1.258 & -1.270 & 1.087(-1) & 2.543(-2) & 1.008 \\
2.2 & 2.915 & 3.414(-1) & 1.260 & -1.269 & 1.103(-1) & 2.533(-2) & 1.009 \\
2.0 & 2.906 & 3.447(-1) & 1.265 & -1.268 & 1.116(-1) & 2.532(-2) & 1.010 \\
\\[-4truemm]
\tableline
\end{tabular}
\end{center}
\tablenum{2}
\caption{Stationary sequence of the equal mass IDR(NS--NS) system 
with $N=0.5$.}
\label{irdrn05.tab}
\end{table}
%


%
\begin{table}
\begin{center}
\begin{tabular}{cccccccccccccc}
\tableline
\tableline
\\[-3truemm]
$\tilde{d}$ & $\bar{d}_G$ & $\bar{\Omega}$ & $\bar{J}$ &
$\bar{E}$ & $T/|W|$ & $\VC$ & $\bar{R}$ \\
\\[-4truemm]
\tableline
\\[-4truemm]
4.0 & 4.141 & 1.949(-1) & 1.440 & -1.190 & 7.362(-2) & 1.384(-2) & 1.001 \\
3.8 & 3.957 & 2.087(-1) & 1.409 & -1.196 & 7.662(-2) & 1.370(-2) & 1.001 \\
3.6 & 3.778 & 2.239(-1) & 1.378 & -1.201 & 7.982(-2) & 1.354(-2) & 1.001 \\
3.4 & 3.603 & 2.407(-1) & 1.347 & -1.208 & 8.325(-2) & 1.337(-2) & 1.001 \\
3.2 & 3.436 & 2.589(-1) & 1.318 & -1.214 & 8.691(-2) & 1.319(-2) & 1.002 \\
3.0 & 3.278 & 2.784(-1) & 1.291 & -1.220 & 9.079(-2) & 1.297(-2) & 1.002 \\
2.8 & 3.135 & 2.986(-1) & 1.267 & -1.226 & 9.481(-2) & 1.271(-2) & 1.003 \\
2.6 & 3.012 & 3.185(-1) & 1.248 & -1.231 & 9.890(-2) & 1.246(-2) & 1.004 \\
2.4 & 2.917 & 3.360(-1) & 1.237 & -1.234 & 1.028(-1) & 1.230(-2) & 1.005 \\
2.3 & 2.884 & 3.429(-1) & 1.234 & -1.235 & 1.045(-1) & 1.219(-2) & 1.006 \\
2.2 & 2.861 & 3.481(-1) & 1.233 & -1.235 & 1.058(-1) & 1.202(-2) & 1.007 \\
2.1 & 2.848 & 3.510(-1) & 1.233 & -1.235 & 1.066(-1) & 1.187(-2) & 1.007 \\
2.0 & 2.847 & 3.515(-1) & 1.233 & -1.235 & 1.068(-1) & 1.175(-2) & 1.008 \\
\\[-4truemm]
\tableline
\end{tabular}
\end{center}
\tablenum{3}
\caption{Stationary sequence of the equal mass IDR(NS--NS) system 
with $N=0.7$.}
\label{irdrn07.tab}
\end{table}
\begin{table}
\begin{center}
\begin{tabular}{cccccccccccccc}
\tableline
\tableline
\\[-3truemm]
$\tilde{d}$ & $\bar{d}_G$ & $\bar{\Omega}$ & $\bar{J}$ &
$\bar{E}$ & $T/|W|$ & $\VC$ & $\bar{R}$ \\
\\[-4truemm]
\tableline
\\[-4truemm]
4.0 & 4.119 & 1.962(-1) & 1.435 & -1.121 & 6.950(-2) & 5.000(-3) & 9.993(-1) \\
3.8 & 3.934 & 2.103(-1) & 1.403 & -1.127 & 7.235(-2) & 4.911(-3) & 9.994(-1) \\
3.6 & 3.753 & 2.259(-1) & 1.371 & -1.133 & 7.541(-2) & 4.814(-3) & 9.996(-1) \\
3.4 & 3.576 & 2.430(-1) & 1.340 & -1.139 & 7.869(-2) & 4.709(-3) & 9.998(-1) \\
3.2 & 3.405 & 2.618(-1) & 1.309 & -1.146 & 8.218(-2) & 4.574(-3) & 1.000 \\
3.0 & 3.244 & 2.820(-1) & 1.280 & -1.153 & 8.585(-2) & 4.416(-3) & 1.001 \\
2.8 & 3.095 & 3.032(-1) & 1.253 & -1.160 & 8.965(-2) & 4.216(-3) & 1.001 \\
2.6 & 2.966 & 3.241(-1) & 1.230 & -1.166 & 9.341(-2) & 3.946(-3) & 1.002 \\
2.4 & 2.864 & 3.427(-1) & 1.213 & -1.171 & 9.683(-2) & 3.666(-3) & 1.004 \\
2.2 & 2.801 & 3.555(-1) & 1.204 & -1.1736 &9.931(-2) & 3.423(-3) & 1.006 \\
2.0 & 2.785 & 3.589(-1) & 1.201 & -1.1744 &9.988(-2) & 3.011(-3) & 1.007 \\
\\[-4truemm]
\tableline
\end{tabular}
\end{center}
\tablenum{4}
\caption{Stationary sequence of the equal mass IDR(NS--NS) system 
with $N=1$.}
\label{irdrn1.tab}
\end{table}
%


%
\begin{table}
\begin{center}
\begin{tabular}{cccccccccccccc}
\tableline
\tableline
\\[-3truemm]
$\tilde{d}$ & $\bar{d}_G$ & $\bar{\Omega}$ & $\bar{J}$ &
$\bar{E}$ & $T/|W|$ & $\VC$ & $\bar{R}$ \\
\\[-4truemm]
\tableline
\\[-4truemm]
4.0 &4.097 &1.976(-1) &1.430 &-9.801(-1) & 6.207(-2) & 1.014(-3) & 9.981(-1) \\
3.8 &3.910 &2.120(-1) &1.397 &-9.859(-1) & 6.469(-2) & 9.685(-4) & 9.983(-1) \\
3.6 &3.726 &2.279(-1) &1.364 &-9.921(-1) & 6.748(-2) & 8.899(-4) & 9.985(-1) \\
3.4 &3.547 &2.455(-1) &1.331 &-9.989(-1) & 7.047(-2) & 7.928(-4) & 9.987(-1) \\
3.2 & 3.373 & 2.648(-1) & 1.299 & -1.006 & 7.365(-2) & 6.723(-4) & 9.991(-1) \\
3.0 & 3.207 & 2.858(-1) & 1.267 & -1.014 & 7.698(-2) & 5.180(-4) & 9.996(-1) \\
2.8 & 3.053 & 3.080(-1) & 1.237 & -1.021 & 8.040(-2) & 3.162(-4) & 1.000 \\
2.6 & 2.916 & 3.302(-1) & 1.211 & -1.029 & 8.373(-2) & 3.760(-5) & 1.002 \\
2.4 & 2.808 & 3.499(-1) & 1.189 & -1.035 & 8.660(-2) & 3.934(-4) & 1.003 \\
2.2 & 2.740 & 3.632(-1) & 1.175 & -1.039 & 8.844(-2) & 9.029(-4) & 1.005 \\
2.0 & 2.726 & 3.658(-1) & 1.170 & -1.041 & 8.858(-2) & 1.287(-3) & 1.006 \\
\\[-4truemm]
\tableline
\end{tabular}
\end{center}
\tablenum{5}
\caption{Stationary sequence of the equal mass IDR(NS--NS) system 
with $N=1.5$.}
\label{irdrn15.tab}
\end{table}
\begin{table}
\begin{center}
\begin{tabular}{cccccccccccccc}
\tableline
\tableline
\\[-3truemm]
$N$ & comment & $r_d$ & $\bar{\Omega}(r_d)$ & $r_c$ & $\bar{\Omega}(r_c)$ & \\
\\[-4truemm]
\tableline
\\[-4truemm]
0   & present & 3.18$\sim$3.28 & 0.283$\sim$0.299 & 3.13  & 0.317  &  \\
    & Paper II& 3.19$\sim$3.28 & 0.282$\sim$0.297 & 3.13  & 0.315  &  \\
    & LRS2    & 3.037          & 0.3191           & 2.842 & 0.3725 &  \\
0.5 & present & 2.94$\sim$3.01 & 0.322$\sim$0.337 & 2.91  & 0.345  &  \\
    & LRS2    & 2.759          & 0.3681           & 2.623 & 0.4091 &  \\
1   & present &  ---           &   ---            & 2.78  & 0.359  &  \\
    & LRS2    & 2.491          & 0.4285           & 2.437 & 0.4466 &  \\
1.5 & present &  ---           &   ---            & 2.73  & 0.366  &  \\
    & LRS2    &  ---           &   ---            & 2.291 & 0.4819 &  \\
\\[-4truemm]
\tableline
\end{tabular}
\end{center}
\tablenum{6}
\caption{Values of $r_c$ and $r_d$, and corresponding $\bar{\Omega}$ 
computed by using the present method, that described in Paper II 
($N=0$ case),  and the ellipsoidal approximation (LRS2) for the equal 
mass IDR binary system.}
\label{ISCO.tab}
\end{table}
%
%

%
%
\begin{table}
\begin{center}
\begin{tabular}{cccccccccccccc}
\tableline
\tableline
\\[-3truemm]
$\tilde{d}$ & $\bar{d}_G$ & $\bar{\Omega}$ & $\bar{J}$ &
$\bar{E}$ & $T/|W|$ & $\VC$ & $\bar{R}$ \\
\\[-4truemm]
\tableline
\\[-4truemm]
4.0 &4.188 &1.913(-1) &1.450 &-7.180(-1) & 1.429(-1) & 5.789(-3) & 9.994(-1) \\
3.8 &4.010 &2.042(-1) &1.419 &-7.231(-1) & 1.476(-1) & 5.771(-3) & 9.994(-1) \\
3.6 &3.838 &2.184(-1) &1.390 &-7.285(-1) & 1.525(-1) & 5.755(-3) & 9.994(-1) \\
3.4 &3.671 &2.337(-1) &1.361 &-7.341(-1) & 1.577(-1) & 5.765(-3) & 9.994(-1) \\
3.2 &3.512 &2.501(-1) &1.334 &-7.399(-1) & 1.630(-1) & 5.818(-3) & 9.994(-1) \\
3.0 &3.365 &2.673(-1) &1.310 &-7.453(-1) & 1.686(-1) & 5.837(-3) & 9.994(-1) \\
2.8 &3.232 &2.849(-1) &1.289 &-7.501(-1) & 1.746(-1) & 5.841(-3) & 9.994(-1) \\
2.6 &3.119 &3.020(-1) &1.275 &-7.538(-1) & 1.807(-1) & 6.075(-3) & 9.994(-1) \\
2.4 &3.031 &3.173(-1) &1.269 &-7.555(-1) & 1.870(-1) & 6.457(-3) & 9.994(-1) \\
2.2 &2.977 &3.288(-1) &1.274 &-7.541(-1) & 1.933(-1) & 6.899(-3) & 9.994(-1) \\
2.0 &2.964 &3.337(-1) &1.288 &-7.501(-1) & 1.980(-1) & 7.425(-3) & 9.994(-1) \\
\\[-4truemm]
\tableline
\end{tabular}
\end{center}
\tablenum{7}
\caption{Stationary sequence of the IRR(BH--NS) system with 
$M_{\ns}/M_{\bh}=1$ and $N=0$.  $\tilde{d}$ is the separation of a point 
source (BH) and the geometrical center of gaseous component star (NS) 
which is defined as $\tilde{d} = ({d}_{\bh}+d_{\ns})/R_0$, where 
$R_0$ is the geometrical radius of the star along the $x$ axis, 
that is $R_0=(R_{\ouou}-R_{\inin})/2$ (see equations (\ref{gdis}) and 
(\ref{normq})).  Other quantities are defined in equations (\ref{pqnorm}), 
(\ref{virial}) and (\ref{rnorm})}
\label{irrreqmn0.tab}
\end{table}
%
%


%
\begin{table}
\begin{center}
\begin{tabular}{cccccccccccccc}
\tableline
\tableline
\\[-3truemm]
$\tilde{d}$ & $\bar{d}_G$ & $\bar{\Omega}$ & $\bar{J}$ &
$\bar{E}$ & $T/|W|$ & $\VC$ & $\bar{R}$ \\
\\[-4truemm]
\tableline
\\[-4truemm]
4.0 & 4.159 & 1.932(-1) & 1.443 & -6.755(-1) & 1.322(-1) & 2.408(-2) & 1.004 \\
3.8 & 3.974 & 2.068(-1) & 1.412 & -6.809(-1) & 1.368(-1) & 2.376(-2) & 1.004 \\
3.6 & 3.794 & 2.219(-1) & 1.380 & -6.868(-1) & 1.416(-1) & 2.341(-2) & 1.004 \\
3.4 & 3.618 & 2.384(-1) & 1.349 & -6.930(-1) & 1.467(-1) & 2.304(-2) & 1.004 \\
3.2 & 3.449 & 2.565(-1) & 1.318 & -6.995(-1) & 1.521(-1) & 2.266(-2) & 1.005 \\
3.0 & 3.288 & 2.759(-1) & 1.289 & -7.061(-1) & 1.577(-1) & 2.223(-2) & 1.005 \\
2.8 & 3.140 & 2.962(-1) & 1.263 & -7.126(-1) & 1.635(-1) & 2.181(-2) & 1.006 \\
2.6 & 3.010 & 3.166(-1) & 1.241 & -7.184(-1) & 1.694(-1) & 2.145(-2) & 1.006 \\
2.4 & 2.904 & 3.354(-1) & 1.226 & -7.227(-1) & 1.752(-1) & 2.108(-2) & 1.008 \\
2.2 & 2.834 & 3.495(-1) & 1.219 & -7.248(-1) & 1.801(-1) & 2.071(-2) & 1.009 \\
\\[-4truemm]
\tableline
\end{tabular}
\end{center}
\tablenum{8}
\caption{Stationary sequence of the IRR(BH--NS) system with 
$M_{\ns}/M_{\bh}=1$ and $N=0.5$.}
\label{irrreqmn05.tab}
\end{table}
\begin{table}
\begin{center}
\begin{tabular}{cccccccccccccc}
\tableline
\tableline
\\[-3truemm]
$\tilde{d}$ & $\bar{d}_G$ & $\bar{\Omega}$ & $\bar{J}$ &
$\bar{E}$ & $T/|W|$ & $\VC$ & $\bar{R}$ \\
\\[-4truemm]
\tableline
\\[-4truemm]
4.0 &4.114 &1.962(-1) &1.434 &-6.216(-1) & 1.221(-1) & 4.394(-3) & 9.993(-1) \\
3.8 &3.927 &2.103(-1) &1.402 &-6.273(-1) & 1.265(-1) & 4.299(-3) & 9.994(-1) \\
3.6 &3.744 &2.261(-1) &1.369 &-6.335(-1) & 1.312(-1) & 4.193(-3) & 9.996(-1) \\
3.4 &3.565 &2.434(-1) &1.336 &-6.401(-1) & 1.361(-1) & 4.073(-3) & 9.998(-1) \\
3.2 & 3.391 & 2.625(-1) & 1.304 & -6.471(-1) & 1.413(-1) & 3.932(-3) & 1.000 \\
3.0 & 3.225 & 2.833(-1) & 1.273 & -6.545(-1) & 1.467(-1) & 3.772(-3) & 1.001 \\
2.8 & 3.070 & 3.053(-1) & 1.243 & -6.621(-1) & 1.523(-1) & 3.577(-3) & 1.001 \\
2.6 & 2.931 & 3.277(-1) & 1.217 & -6.693(-1) & 1.577(-1) & 3.310(-3) & 1.003 \\
2.4 & 2.818 & 3.482(-1) & 1.196 & -6.755(-1) & 1.628(-1) & 3.053(-3) & 1.004 \\
\\[-4truemm]
\tableline
\end{tabular}
\end{center}
\tablenum{9}
\caption{Stationary sequence of the IRR(BH--NS) system with 
$M_{\ns}/M_{\bh}=1$ and $N=1$.}
\label{irrreqmn1.tab}
\end{table}
%


%
\begin{table}
\begin{center}
\begin{tabular}{cccccccccccccc}
\tableline
\tableline
\\[-3truemm]
$\tilde{d}$ & $\bar{d}_G$ & $\bar{\Omega}$ & $\bar{J}$ &
$\bar{E}$ & $T/|W|$ & $\VC$ & $\bar{R}$ \\
\\[-4truemm]
\tableline
\\[-4truemm]
4.0 &3.464 &2.207(-1) &2.420 &-8.270(-1) &2.180(-1) & 5.414(-3) & 9.994(-1) \\
3.8 &3.340 &2.333(-1) &2.379 &-8.350(-1) &2.230(-1) & 5.466(-3) & 9.994(-1) \\
3.6 &3.223 &2.465(-1) &2.342 &-8.429(-1) &2.281(-1) & 5.490(-3) & 9.994(-1) \\
3.4 &3.114 &2.601(-1) &2.307 &-8.504(-1) &2.334(-1) & 5.621(-3) & 9.994(-1) \\
3.2 &3.015 &2.737(-1) &2.278 &-8.572(-1) &2.388(-1) & 5.861(-3) & 9.994(-1) \\
3.0 &2.931 &2.866(-1) &2.255 &-8.627(-1) &2.442(-1) & 6.117(-3) & 9.994(-1) \\
2.8 &2.863 &2.982(-1) &2.242 &-8.662(-1) &2.496(-1) & 6.563(-3) & 9.994(-1) \\
2.6 &2.816 &3.072(-1) &2.240 &-8.666(-1) &2.549(-1) & 7.053(-3) & 9.994(-1) \\
2.4 &2.797 &3.123(-1) &2.251 &-8.636(-1) &2.596(-1) & 7.667(-3) & 9.994(-1) \\
2.2 &2.812 &3.105(-1) &2.262 &-8.608(-1) &2.597(-1) & 8.284(-3) & 9.994(-1) \\
\\[-4truemm]
\tableline
\end{tabular}
\end{center}
\tablenum{10}
\caption{Stationary sequence of the IRR(BH--NS) system with 
$M_{\ns}/M_{\bh}=0.5$ and $N=0$.}
\label{irrrm2n0.tab}
\end{table}
\begin{table}
\begin{center}
\begin{tabular}{cccccccccccccc}
\tableline
\tableline
\\[-3truemm]
$\tilde{d}$ & $\bar{d}_G$ & $\bar{\Omega}$ & $\bar{J}$ &
$\bar{E}$ & $T/|W|$ & $\VC$ & $\bar{R}$ \\
\\[-4truemm]
\tableline
\\[-4truemm]
4.0 & 3.407 & 2.259(-1) & 2.397 & -7.877(-1) & 2.057(-1) & 1.881(-2) & 1.005 \\
3.8 & 3.275 & 2.400(-1) & 2.352 & -7.968(-1) & 2.109(-1) & 1.842(-2) & 1.005 \\
3.6 & 3.148 & 2.549(-1) & 2.308 & -8.062(-1) & 2.162(-1) & 1.801(-2) & 1.005 \\
3.4 & 3.028 & 2.705(-1) & 2.267 & -8.155(-1) & 2.215(-1) & 1.761(-2) & 1.006 \\
3.2 & 2.918 & 2.864(-1) & 2.230 & -8.245(-1) & 2.270(-1) & 1.726(-2) & 1.006 \\
3.0 & 2.821 & 3.020(-1) & 2.198 & -8.325(-1) & 2.324(-1) & 1.690(-2) & 1.007 \\
2.8 & 2.741 & 3.162(-1) & 2.174 & -8.388(-1) & 2.376(-1) & 1.651(-2) & 1.009 \\
2.6 & 2.685 & 3.272(-1) & 2.160 & -8.428(-1) & 2.420(-1) & 1.615(-2) & 1.011 \\
2.4 & 2.661 & 3.324(-1) & 2.156 & -8.439(-1) & 2.444(-1) & 1.602(-2) & 1.012 \\
\\[-4truemm]
\tableline
\end{tabular}
\end{center}
\tablenum{11}
\caption{Stationary sequence of the IRR(BH--NS) system with 
$M_{\ns}/M_{\bh}=0.5$ and $N=0.5$.}
\label{irrrm2n05.tab}
\end{table}
%


%
\begin{table}
\begin{center}
\begin{tabular}{cccccccccccccc}
\tableline
\tableline
\\[-3truemm]
$\tilde{d}$ & $\bar{d}_G$ & $\bar{\Omega}$ & $\bar{J}$ &
$\bar{E}$ & $T/|W|$ & $\VC$ & $\bar{R}$ \\
\\[-4truemm]
\tableline
\\[-4truemm]
4.0 & 3.353 & 2.311(-1) & 2.374 & -7.366(-1) & 1.934(-1) & 2.959(-3) & 1.000 \\
3.8 & 3.217 & 2.461(-1) & 2.327 & -7.466(-1) & 1.985(-1) & 2.822(-3) & 1.000 \\
3.6 & 3.085 & 2.621(-1) & 2.280 & -7.568(-1) & 2.037(-1) & 2.658(-3) & 1.001 \\
3.4 & 2.961 & 2.790(-1) & 2.235 & -7.674(-1) & 2.090(-1) & 2.450(-3) & 1.002 \\
3.2 & 2.845 & 2.964(-1) & 2.192 & -7.780(-1) & 2.143(-1) & 2.201(-3) & 1.003 \\
3.0 & 2.742 & 3.135(-1) & 2.155 & -7.878(-1) & 2.194(-1) & 1.940(-3) & 1.004 \\
2.8 & 2.658 & 3.289(-1) & 2.125 & -7.963(-1) & 2.240(-1) & 1.707(-3) & 1.006 \\
2.6 & 2.604 & 3.397(-1) & 2.105 & -8.019(-1) & 2.272(-1) & 1.322(-3) & 1.008 \\
\\[-4truemm]
\tableline
\end{tabular}
\end{center}
\tablenum{12}
\caption{Stationary sequence of the IRR(BH--NS) system with 
$M_{\ns}/M_{\bh}=0.5$ and $N=1$.}
\label{irrrm2n1.tab}
\end{table}
\begin{table}
\begin{center}
\begin{tabular}{cccccccccccccc}
\tableline
\tableline
\\[-3truemm]
$\tilde{d}$ & $\bar{d}_G$ & $\bar{\Omega}$ & $\bar{J}$ &
$\bar{E}$ & $T/|W|$ & $\VC$ & $\bar{R}$ \\
\\[-4truemm]
\tableline
\\[-4truemm]
6.0 & 3.129 & 2.200(-1) & 7.836 & -1.340 & 3.566(-1) & 4.653(-3) & 9.994(-1) \\
5.8 & 3.061 & 2.275(-1) & 7.753 & -1.356 & 3.592(-1) & 4.726(-3) & 9.994(-1) \\
5.6 & 2.994 & 2.352(-1) & 7.672 & -1.372 & 3.617(-1) & 4.860(-3) & 9.994(-1) \\
5.4 & 2.931 & 2.430(-1) & 7.595 & -1.388 & 3.641(-1) & 5.044(-3) & 9.994(-1) \\
5.2 & 2.872 & 2.507(-1) & 7.523 & -1.404 & 3.666(-1) & 5.239(-3) & 9.994(-1) \\
5.0 & 2.816 & 2.584(-1) & 7.456 & -1.419 & 3.690(-1) & 5.480(-3) & 9.994(-1) \\
4.8 & 2.765 & 2.658(-1) & 7.395 & -1.433 & 3.714(-1) & 5.816(-3) & 9.994(-1) \\
4.6 & 2.719 & 2.729(-1) & 7.341 & -1.445 & 3.737(-1) & 6.236(-3) & 9.994(-1) \\
4.4 & 2.679 & 2.794(-1) & 7.297 & -1.456 & 3.760(-1) & 6.732(-3) & 9.994(-1) \\
4.2 & 2.646 & 2.851(-1) & 7.263 & -1.464 & 3.782(-1) & 7.363(-3) & 9.994(-1) \\
4.0 & 2.621 & 2.897(-1) & 7.242 & -1.470 & 3.803(-1) & 8.102(-3) & 9.994(-1) \\
3.8 & 2.605 & 2.929(-1) & 7.235 & -1.471 & 3.822(-1) & 9.019(-3) & 9.994(-1) \\
3.6 & 2.599 & 2.944(-1) & 7.239 & -1.470 & 3.835(-1) & 1.026(-2) & 9.994(-1) \\
3.4 & 2.602 & 2.942(-1) & 7.247 & -1.468 & 3.837(-1) & 1.135(-2) & 9.994(-1) \\
\\[-4truemm]
\tableline
\end{tabular}
\end{center}
\tablenum{13}
\caption{Stationary sequence of the IRR(BH--NS) system with 
$M_{\ns}/M_{\bh}=0.1$ and $N=0$.}
\label{irrrm10n0.tab}
\end{table}
%


%
\begin{table}
\begin{center}
\begin{tabular}{cccccccccccccc}
\tableline
\tableline
\\[-3truemm]
$\tilde{d}$ & $\bar{d}_G$ & $\bar{\Omega}$ & $\bar{J}$ &
$\bar{E}$ & $T/|W|$ & $\VC$ & $\bar{R}$ \\
\\[-4truemm]
\tableline
\\[-4truemm]
6.0 & 3.060 & 2.273(-1) & 7.744 & -1.314 & 3.470(-1) & 8.369(-3) & 1.005 \\
5.8 & 2.986 & 2.359(-1) & 7.652 & -1.332 & 3.497(-1) & 8.115(-3) & 1.006 \\
5.6 & 2.914 & 2.447(-1) & 7.562 & -1.351 & 3.525(-1) & 7.864(-3) & 1.006 \\
5.4 & 2.845 & 2.538(-1) & 7.475 & -1.370 & 3.552(-1) & 7.633(-3) & 1.006 \\
5.2 & 2.780 & 2.629(-1) & 7.391 & -1.389 & 3.580(-1) & 7.408(-3) & 1.007 \\
5.0 & 2.718 & 2.722(-1) & 7.312 & -1.407 & 3.607(-1) & 7.147(-3) & 1.008 \\
4.8 & 2.660 & 2.812(-1) & 7.239 & -1.424 & 3.633(-1) & 6.836(-3) & 1.008 \\
4.6 & 2.608 & 2.900(-1) & 7.173 & -1.441 & 3.659(-1) & 6.504(-3) & 1.009 \\
4.4 & 2.561 & 2.981(-1) & 7.115 & -1.455 & 3.683(-1) & 6.123(-3) & 1.011 \\
4.2 & 2.523 & 3.053(-1) & 7.067 & -1.468 & 3.704(-1) & 5.676(-3) & 1.012 \\
4.0 & 2.493 & 3.111(-1) & 7.031 & -1.478 & 3.723(-1) & 5.201(-3) & 1.014 \\
3.8 & 2.474 & 3.150(-1) & 7.009 & -1.484 & 3.735(-1) & 4.760(-3) & 1.015 \\
3.7 & 2.469 & 3.161(-1) & 7.002 & -1.485 & 3.738(-1) & 4.578(-3) & 1.015 \\
\\[-4truemm]
\tableline
\end{tabular}
\end{center}
\tablenum{14}
\caption{Stationary sequence of the IRR(BH--NS) system with 
$M_{\ns}/M_{\bh}=0.1$ and $N=0.5$.}
\label{irrrm10n05.tab}
\end{table}
\begin{table}
\begin{center}
\begin{tabular}{cccccccccccccc}
\tableline
\tableline
\\[-3truemm]
$\tilde{d}$ & $\bar{d}_G$ & $\bar{\Omega}$ & $\bar{J}$ &
$\bar{E}$ & $T/|W|$ & $\VC$ & $\bar{R}$ \\
\\[-4truemm]
\tableline
\\[-4truemm]
6.0 & 3.000 & 2.339(-1) & 7.663 & -1.275 & 3.363(-1) & 2.059(-4) & 1.001 \\
5.8 & 2.924 & 2.432(-1) & 7.565 & -1.295 & 3.392(-1) & 8.263(-5) & 1.002 \\
5.6 & 2.849 & 2.528(-1) & 7.469 & -1.315 & 3.421(-1) & 6.050(-5) & 1.002 \\
5.4 & 2.778 & 2.627(-1) & 7.376 & -1.336 & 3.450(-1) & 2.241(-4) & 1.003 \\
5.2 & 2.709 & 2.729(-1) & 7.285 & -1.357 & 3.478(-1) & 4.043(-4) & 1.004 \\
5.0 & 2.644 & 2.831(-1) & 7.198 & -1.378 & 3.506(-1) & 5.893(-4) & 1.005 \\
4.8 & 2.583 & 2.932(-1) & 7.117 & -1.398 & 3.532(-1) & 7.808(-4) & 1.006 \\
4.6 & 2.529 & 3.029(-1) & 7.044 & -1.417 & 3.557(-1) & 9.980(-4) & 1.007 \\
4.4 & 2.481 & 3.117(-1) & 6.980 & -1.434 & 3.579(-1) & 1.278(-3) & 1.009 \\
4.2 & 2.445 & 3.188(-1) & 6.930 & -1.448 & 3.597(-1) & 1.648(-3) & 1.010 \\
4.0 & 2.421 & 3.237(-1) & 6.895 & -1.458 & 3.607(-1) & 2.245(-3) & 1.012 \\
\\[-4truemm]
\tableline
\end{tabular}
\end{center}
\tablenum{15}
\caption{Stationary sequence of the IRR(BH--NS) system with 
$M_{\ns}/M_{\bh}=0.1$ and $N=1$.}
\label{irrrm10n1.tab}
\end{table}

\clearpage

%
%
%
%
%
{}

\clearpage

%
%
%
\figcaption{
Physical quantities for a stationary sequence of the compressible IDR binary 
system with $N=1.5$ polytropes.  (a) Total angular momentum
as a function of a binary separation $\Jseq$. (b) Total energy as a 
function of a binary separation $\Eseq$. (c) Orbital angular velocity 
as a function of the total angular momentum $\Oseq$.  Solid curves and 
dots show the results of LRS1 and our present results of irrotational 
binary stars, respectively.  See text about the normalization factors for 
each quantity.
\label{irdrn15.gra}}
\figcaption{
Physical quantities for the IDR sequences with several 
polytropic indices.  (a) Total angular momentum as a function of a binary
separation $\Jseq$.  (b) Orbital angular velocity as a function of 
the total angular momentum $\Oseq$.  Different curves correspond to 
different polytropic indices:
$N=0$ (dash dotted line), $N=0.5$ (dotted line), $N=0.7$ (long dashed line), 
$N=1$ (solid line) and $N=1.5$ (short dashed line).  Dots show the results 
for the $N=0$ sequence computed by using a different scheme described in 
Paper II.  
\label{irdrall.gra}}
%
%
%
%
\figcaption{Distributions of physical quantities for the IDR binary systems 
with $N=0.5$ polytropes.  
(a) Contours of the density in the equatorial $x$-$y$ plane.  
(b) Contours of the density in the meridional $x$-$z$ plane.
(c) Contours of the density in the plane which is parallel to the
$y$-$z$ plane and intersects with the geometrical center of the star.  
(d) Contours of the velocity potential in the equatorial $x$-$y$ 
plane.  The difference between two subsequent contours for each quantity is 
$1/10$ of the difference between maximum to minimum value.  ${\tilde X}$, 
${\tilde Y}$ and ${\tilde Z}$ are the Cartesian coordinates parallel to the 
$xyz$--coordinates in equation (\ref{sphcor}) whose origin is at the 
geometrical center of the star $r=0$ and normalized as the equation 
(\ref{normq}).  This configuration corresponds to that of the contact phase 
which is dynamically unstable.
\label{irdrn05.con}}
\figcaption{Same as Figure~\protect\ref{irdrn05.con} but for the IDR binary 
system with $N=1.0$ polytropes.  This configuration is dynamically stable.  
\label{irdrn1.con}}
\figcaption{Same as Figure~\protect\ref{irdrn05.con} but for the IDR binary 
system with $N=1.5$ polytropes.  This configuration is also dynamically 
stable.
\label{irdrn15.con}}
\figcaption{Distributions of physical quantities for the IDR binary system 
with $N=0.5$ polytropes.  
Panels (a)--(d) show the same quantities as in Figure~\ref{irdrn05.con}.  
(e) Contours of the velocity potential in the plane which is 
parallel to the $y$-$z$ plane and intersects with the geometrical center 
of the star.  
Contours are drawn in the same way as those in Figure~ \ref{irdrn05.con}.  
(f) The velocity field seen from the rotational frame 
in the equatorial $x$-$y$ plane ${\bf u}$.  The length of arrows is 
normalized by equation~(\ref{cpnorm}).  This configuration corresponds to 
that for $\tilde{d}=2.4$ which is near the turning point of $\Jseq$ or 
$\Eseq$ of the stationary sequence.  
\label{irdrn05.con2}}
\figcaption{Surface shape of a component star of the IDR binary system 
with $N=0.5$ polytrope and $\tilde{d}=2.4$.  This configuration corresponds 
to the model shown in Figure~\ref{irdrn05.con2}.  
\label{irdrn05.surf}}
\figcaption{The velocity profiles of the IDR binary systems along the 
$x$-axis. (a) At the contact phase for $N=0$ (dash dotted line), $0.5$ 
(dotted line), $1$ (solid line) and $1.5$ (short dashed line) polytropes.  
(b) At the distance of $\tilde{d}=2.4$ for an $N=0.5$ polytrope.  
\label{irdrvec.prof}}
%
%
%
%
\figcaption{
Comparison of the results obtained by using the present method, that of 
LRS1 and that of Paper II for the equal mass ($\ratmas=1$) and 
incompressible ($N=0$) IRR binary systems.  Panels (a), (b) and (c) 
are the same as in Figure~\ref{irdrn15.gra}.   Dashed and solid curves 
show the results of LRS1 and our present results for irrotational binary 
stars, respectively.   Dots show the results for the $N=0$ sequence 
computed by using a different scheme described in Paper II.  
\label{irrreqmn0.gra}}
\figcaption{Same as Figure~\protect\ref{irrreqmn0.gra}
but for the mass ratio $\ratmas=0.1$
\label{irrrm10n0.gra}}
\figcaption{Comparison of the results obtained by using the present 
method and that of LRS1 for the equal mass ($\ratmas=1$) and $N=1$ IRR 
binary systems.  Panels~ (a), (b) and (c) are the same as in 
Figure~\ref{irdrn15.gra}.  Solid curves and dots correspond to 
the results of LRS1 and the present results, respectively.  
\label{irrreqmn1.gra}}
\figcaption{Same as Figure~\protect\ref{irrreqmn1.gra}
but for the mass ratio $\ratmas=0.1$
\label{irrrm10n1.gra}}
\figcaption{Same as Figure~\protect\ref{irdrall.gra} 
but for the IRR binary systems with the mass ratio $\ratmas=0.5$.  
Solid lines correspond to the stationary sequences with polytropic indices 
$N=0$ (dash dotted line), $N=0.1$ (dotted line), $N=0.3$ (long dashed line),
$N=0.5$ (solid line) and $N=1$ (short dashed line), respectively.  
Dots show the results with $N=0$ sequence computed by using a different 
scheme described in Paper II.  
\label{irrrm2all.gra}}
%
%
%
%
\figcaption{Distributions of physical quantities of the IRR binary system 
with $\ratmas=0.5$ and $N=0.5$.  Panels (a)--(d) and (f) are the same as 
in Figure~ \ref{irdrn05.con2}.  (e) Contours of the velocity potential in 
through $\varphi=\pi/16$ to $\varphi=17\pi/16$ plane.  The length of 
arrows is normalized as equation (\ref{normq}).  This configuration 
corresponds to the last entry of Table \ref{irrrm2n05.tab} which is 
dynamically stable.  
\label{irrrm2n05.con}}
\figcaption{Same as Figure~\ref{irdrn05.con} but for the IRR binary system 
with $N=1$ and $\ratmas=0.5$.  This configuration corresponds to the last 
entry of the Table~\ref{irrrm2n1.tab} which is dynamically stable.
\label{irrrm2n1.con}}
\figcaption{Surface shape of the gaseous component of the IRR binary system 
with $N=0.5$ and $\ratmas=0.1$ polytropes.  This configuration corresponds 
to the last entry 
of Table~\ref{irrrm10n05.tab}, which is dynamically stable.
\label{irrrm10n05.surf}}
\figcaption{Schematic summary of our results and those of LRS1 for the
irrotational binary systems.  The length $d_G$ denotes the separation of 
two centers of mass of two stars. Three distances $r_d$, $r_c$ and $r_R$ 
denote the hydrodynamical stability limit and the contact limit and the 
Roche--Riemann limit, respectively.  
\label{figsummary}}


\begin{thebibliography}{}
%
%
\bibitem[Abramovici et al. 1992]{ab92}
  Abramovici A. et al., 1992, Science, 256, 325
\bibitem[Aizenman 1968]{ai68}
  Aizenman M. L., 1968, \apj, 153, 511
\bibitem[Asada 1998]{asa98}
  Asada .H, 1998, \prd, in press, (gr-qc/9804003)
\bibitem[Baumgarte et al. 1997a]{ba97a}
  Baumgarte T. W., Cook G. B., Scheel M. A., Shapiro S. L., 
  Teukolsky S. A., 1997a, \prl, 79, 1182-1185, (gr-qc/9704024)
\bibitem[Baumgarte et al. 1997b]{ba97b}
  Baumgarte T. W., Cook G. B., Scheel M. A., Shapiro S. L., 
  Teukolsky S. A., 1997b, \prl, submitted, (gr-qc/9705023)
\bibitem[Baumgarte et al. 1997c]{ba97c}
  Baumgarte T. W., Cook G. B., Scheel M. A., Shapiro S. L., 
  Teukolsky S. A., 1997c, \prd, submitted, (gr-qc/9709026)
\bibitem[Bildsten \& Cutler 1992]{bc92}
  Bildsten L., Cutler C., 1992, \apj, 400, 175
\bibitem[Bonazzola, Gourgoulhon \& Marck 1997]{bgm97}
  Bonazzola S., Gourgoulhon E., Marck J.-A., 1997, \prd, in press
  (gr-qc/9710031)
\bibitem[Chandrasekhar 1969]{ch69}
  Chandrasekhar S., 1969, Ellipsoidal Figures of Equilibrium,
  Yale University Press, New Haven
\bibitem[Davis et. al. 1994]{da94}
  Davis M., Benz W., Piran T., Thielemann F., 1994, \apj, 431, 742
%
\bibitem[Eriguchi 1990]{eri90}
  Eriguchi Y., 1990, \aap, 229, 457
\bibitem[Eriguchi \& Hachisu 1983]{eh83}
  Eriguchi Y., Hachisu I., 1983, Prog. Theor. Phys., 70, 1534
\bibitem[Eriguchi \& Hachisu 1985]{eh85}
  Eriguchi Y., Hachisu I., 1985, \aap, 142, 256
\bibitem[Eriguchi \& M\"uller 1985]{em85}
  Eriguchi Y., M\"uller E., 1985, \aap, 146, 260
\bibitem[Eriguchi, Futamase \& Hachisu 1990]{efh90}
  Eriguchi Y., Futamase T., Hachisu I., 1990, \aap, 231, 61
\bibitem[Eriguchi, Hachisu \& Sugimoto 1982]{ehs82}
  Eriguchi Y., Hachisu I., Sugimoto D., 1982, Prog. Theor. Phys., 67, 1068
\bibitem[Hachisu 1986]{ha86}
  Hachisu I., 1986, \apjs, 62, 461
\bibitem[Hachisu \& Eriguchi 1984a]{he84a}
  Hachisu I., Eriguchi Y., 1984a, \pasj, 36, 239
\bibitem[Hachisu \& Eriguchi 1984b]{he84b}
  Hachisu I., Eriguchi Y., 1984b, \pasj, 36, 259
%
\bibitem[Kochanek 1992]{ko92} 
  Kochanek C. S., 1992, \apj, 398, 234
\bibitem[Komatsu, Eriguchi \& Hachisu 1989]{keh89}
  Komatsu H., Eriguchi Y.  Hachisu I., 1989, \mnras, 237, 355
%
\bibitem[Lai, Rasio \& Shapiro 1993]{lrs93a} 
  Lai D., Rasio F. A., Shapiro S. L., 1993a, \apj,  406, L63
\bibitem[Lai, Rasio \& Shapiro 1993]{lrs93b} 
  Lai D., Rasio F. A., Shapiro S. L., 1993b, \apjs,  88, 205, LRS1
\bibitem[Lai, Rasio \& Shapiro 1994]{lrs94a} 
  Lai D., Rasio F. A., Shapiro S. L., 1994a, \apj, 420, 811, LRS2
\bibitem[Lai, Rasio \& Shapiro 1994]{lrs94b} 
  Lai D., Rasio F. A., Shapiro S. L., 1994b, \apj, 423, 344
\bibitem[Lombardi, Rasio \& Shapiro 1997]{lrs97} 
  Lombardi J. C., Rasio F. A., Shapiro S. L., 1997, \prd, 56, 3416
%
\bibitem[Lee \& Klu\'zniak 1997]{lk97} 
  Lee W. H., Klu\'zniak W., 1997, in preparation
\bibitem[Mathews, Marronetti \& Wilson 1997]{mmw97}  
  Mathews G .J., Marronetti P., Wilson J.R., 1997, \prd, submitted, 
  (gr-qc/9710140)
\bibitem[Miller 1974]{mi74} 
  Miller B.D., 1974, \apj, 187, 609
\bibitem[Ostriker \& Mark 1968]{om68} 
  Ostriker J. P., Mark J. W-K., 1968, \apj, 151, 1075
\bibitem[Paczy\'nski 1986]{pa86} 
  Paczy\'nski B., 1986, \apj, 308, L43
%
\bibitem[Rasio \& Shapiro 1994]{rs94}
  Rasio F. A., Shapiro S. L., 1994, \apj, 432, 242
\bibitem[Rasio \& Shapiro 1996]{rs96}
  Rasio F. A., Shapiro S. L., 1996, in Compact Stars in Binaries, 
  proceedings of IAU Symposium 165, eds. 
  van Paradijs J., van den Heuvel E. P. J., Kuulkers E.,
  Dordrecht, Kluwer Academic Publishers, (astro-ph/9410085)
%
\bibitem[Ruffert, Rampp \& Janka 1997]{rrj97}
  Ruffert M., Rampp M., Janka H.-T., 1997, \aap , 321, 991
\bibitem[Ruffert et al. (1997)]{rjts97}
  Ruffert M., Janka H.-T., Takahashi K., Schaefer G., 
  1997, \aap , 319, 122
%
\bibitem[Shapiro \& Teukolsky 1983]{st83}
  Shapiro S. L., Teukolsky S. A., 1983, 
  Black Holes, White Dwarfs and Neutron Stars, Wiley, New York
\bibitem[Shibata 1997]{sh97} 
  Shibata M., 1997, \prd, 55, 6019
\bibitem[Shibata 1998]{sh98}
  Shibata M., 1998, \prd, in press, (gr-qc/9803085)
\bibitem[Shibata, Nakamura \& Oohara 1992]{sno92} 
  Shibata M., Nakamura. T., Oohara K., 1992, Prog. Theor. Phys., 88, 1079
\bibitem[Shibata, Oohara \& Nakamura 1997]{son97} 
  Shibata M., Oohara K., Nakamura. T., 1997, Prog. Theor. Phys., 98, in press
\bibitem[Taniguchi \& Nakamura 1996]{tn96}  
  Taniguchi K., Nakamura T., 1996, Prog. Theor. Phys., 96, 693
%
\bibitem[Tassoul 1978]{tas78} Tassoul J.-L., 1978, Theory of Rotating Stars, 
    Princeton Univ. Press, Princeton
\bibitem[Taub 1959]{taub59} 
    Taub A. H., 1959, Arch. Ratl. Mech. Anal., 3, 312 
\bibitem[Teukolsky 1998]{teu98}
  Teukolsky S. A., 1998, \apj, in press, (gr-qc/9803082)
%
\bibitem[Thorne 1994]{th94}
  Thorne K. S., 1994, in ``Relativistic Cosmology" 
  Proceedings of the 8-th Nishinomiya-Yukawa Memorial Symposium, 
  ed. Sasaki M., 
  Universal Academy Press, Tokyo, p. 67
\bibitem[Ury\=u \& Eriguchi 1996]{ue96}  
  Ury\=u K., Eriguchi Y., 1996, \mnras, 282, 653 
\bibitem[Ury\=u \& Eriguchi 1998a]{ue98a}  
  Ury\=u K., Eriguchi Y., 1998a, \mnras, 296, L1, Paper I 
  (astro-ph/9712203) 
\bibitem[Ury\=u \& Eriguchi 1998b]{ue98b}  
  Ury\=u K., Eriguchi Y., 1998b, \mnras, in press, Paper II
\bibitem[Ury\=u \& Eriguchi 1998c]{ue98c}  
  Ury\=u K., Eriguchi Y., 1998c, in preparation
%
\bibitem[Wilson, Mathews \& Marronetti 1996]{wmm96}  
  Wilson J .R., Mathews G .J., Marronetti P., 1996, \prd 54, 1317
\bibitem[Yoshida \& Eriguchi 1997]{ye97}  
  Yoshida S., Eriguchi Y., 1997, \prd 56, 762
\bibitem[Zhuge, Centrella \& McMillan 1994]{zcm94}  
  Zhuge X., Centrella J. M., McMillan S. L. W., 1994, \prd 50, 6247
\bibitem[Zhuge, Centrella \& McMillan 1996]{zcm96}  
  Zhuge X., Centrella J. M., McMillan S. L. W., 1996, \prd 54, 7261
%
\end{thebibliography}
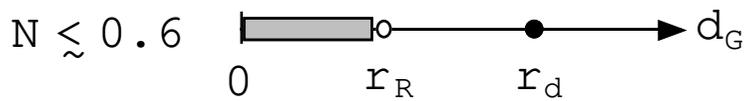
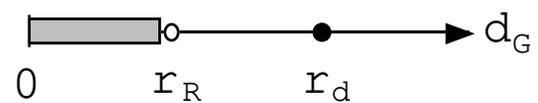
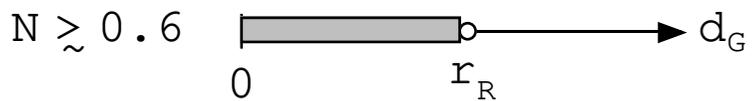
\end{document}